\documentclass[a4paper]{article}

\usepackage[french,english]{babel}
\usepackage[T1]{fontenc}
\usepackage[utf8]{inputenc}
\usepackage{amsmath,amssymb,amsfonts,amsthm}
\usepackage{makeidx}
\usepackage{color}
\usepackage{empheq}
\usepackage{cancel}
\usepackage[top=20mm, bottom=20mm, left=20mm, right=20mm]{geometry}
\usepackage{stmaryrd} 
\usepackage{enumerate}
\usepackage{hyperref} 
\usepackage{booktabs}
\usepackage[all]{xy} 

\usepackage[vcentermath,noautoscale]{youngtab}
\everymath{\Yboxdim10pt}
\everydisplay{\Yboxdim10pt}

\SetSymbolFont{stmry}{bold}{U}{stmry}{m}{n} 

\theoremstyle{remark}

\theoremstyle{definition}

\newcommand{\Ac}{\mathcal{A}}

\newcommand{\Ec}{\mathcal{E}}

\newcommand{\Lc}{\mathcal{L}}

\newcommand{\Mc}{\mathcal{M}}
\newcommand{\Mr}{\mathrm{M}}
\newcommand{\Mb}{\mathbb{M}}
\newcommand{\Mbd}{\dot{\mathbb{M}}}
\newcommand{\Mbdd}{\ddot{\mathbb{M}}}
\newcommand{\Nb}{\mathbb{N}}
\newcommand{\Nbd}{\dot{\mathbb{N}}}
\newcommand{\Nbdd}{\ddot{\mathbb{N}}}
\newcommand{\NNb}{\mathbb{N} \negthinspace \mathbb{N}}
\newcommand{\NNbd}{\dot{\mathbb{N} \negthinspace \mathbb{N}}}
\newcommand{\NNbdd}{\ddot{\mathbb{N} \negthinspace \mathbb{N}}}
\newcommand{\Nr}{\mathrm{N}}
\newcommand{\Nrb}{\overline{\mathrm{N}}}
\newcommand{\Or}{\mathrm{O}}
\newcommand{\Orb}{\overline{\mathrm{O}}}
\newcommand{\Ob}{\mathbb{O}}
\newcommand{\Obd}{\dot{\mathbb{O}}}
\newcommand{\Obdd}{\ddot{\mathbb{O}}}

\newcommand{\Rr}{\mathrm{R}}
\newcommand{\Rb}{\mathbb{R}}
\newcommand{\Rbd}{\dot{\mathbb{R}}}
\newcommand{\Rbdd}{\ddot{\mathbb{R}}}
\newcommand{\Sc}{\mathcal{S}}

\newcommand{\Vc}{\mathcal{V}}

\renewcommand{\d}{\ensuremath{\operatorname{d}\!}}

\newcommand{\derd}[2]{\frac{\delta #1}{\delta #2}}

\newcommand{\derp}[2]{\frac{\partial#1}{\partial#2}}

\newcommand{\drond}{\partial}

\title{\textsf{\textbf{\texorpdfstring{$n+1$}{n+1} formalism of $f(\text{Lovelock})$ gravity}}}
\author{\textsf{Xavier Lachaume}}
\date{\textsf{\normalsize{Summer 2017}}}

\begin{document}
\maketitle

\begin{center}
\textsf{
Institut Denis Poisson \\
Université de Tours - UMR 7013 du CNRS \\
Parc de Grandmont - 37200 Tours - France
}

\texttt{xavier.lachaume@lmpt.univ-tours.fr}
\end{center}

\renewcommand*{\proofname}{$\textstyle\square \scriptstyle\square \scriptscriptstyle\square \diamond$}
\renewcommand{\qedsymbol}{$. \scriptscriptstyle \diamond \square \scriptstyle\square \textstyle\square$}

\vspace{15mm}

\textsc{Abstract:}
In this note we perform the $n+1$ decomposition, or Arnowitt–Deser–Misner (ADM) formulation of $f(\text{Lovelock})$ gravity theory. The hamiltonian form of Lovelock gravity was known since the work of C. Teitelboim and J. Zanelli in 1987, but this result had not yet been extended to $f(\text{Lovelock})$ gravity. Besides, field equations of $f(\text{Lovelock})$ have been recently computed by P. Bueno et al., though without ADM decomposition.

We focus on the non-degenerate case, ie. when the Hessian of $f$ is invertible. Using the same Legendre transform as for $f(\Rr)$ theories, we can identify the partial derivatives of $f$ as scalar fields, and consider the theory as a generalised scalar-tensor theory. We then derive the field equations, and project them along a $n+1$ decomposition. We obtain an original system of constraint equations for $f(\text{Lovelock})$ gravity, as well as dynamical equations. We give explicit formulas for the $f(\Rr, \text{Gauss-Bonnet})$ case.

\vspace{15mm}




\section{Introduction}


As a gravitation theory, General Relativity (GR) has been tested with very high degrees of precision in the weak-field limit: almost all the observations and experiments performed in the solar system and around the earth confirm the accuracy of the predictions of this theory. One can find a review on this topic in \cite{WillCM14}. In stronger gravity regimes, however, this has not yet been achieved. The recent discoveries of gravitational waves, eg. \cite{Abbott16}, promise for the following years interesting measures about the behaviour of matter and gravitation in strong gravitation fields. Thus might be detected what could have escaped to weak-field measures up to now (see \cite{Baker17}, \cite{Chamberlain17}): a modification from GR with higher-degree curvature terms.

Indeed, such terms are negligible when the curvature is small. Nevertheless, there are several clues coming from theoretical physics, observation, or even geometry, that higher-degree curvatures could be interesting to include in the depiction of gravitation. The Gauss-Bonnet curvature, which is a curvature term of order 2, appears naturally in some string theories. It could explain part of the current cosmological observations, without assuming the existence of enigmatic elements such as dark matter or dark energy. And the Lovelock theories, which are a generalisation of this Gauss-Bonnet curvature in higher dimension, are connected in an amazing way with the Gauss-Bonnet-Chern theorem, a fundamental geometry theorem.

An other set of modified gravity theories has raised a renewed interest in the last years, for the same cosmological motivations and because of its simplicity as a toy-model: the $f(\Rr)$ theories (\cite{Vainio17}). Unlike the Lovelock theories, they generate field equations with higher-order derivatives. We could as well have mentioned scalar-tensor, Horndeski (\cite{Hou17}, \cite{Kreisch17}, \cite{Arai17}) or Born-Infeld (\cite{Jimenez17}, \cite{Jana17}) theories.

The natural union between Lovelock and $f(\Rr)$ theories, called -- with originality -- $f($Lovelock$)$ theories, were not studied up to recently, in \cite{Bueno16}. There are still many properties of these theories to investigate. In the present paper, we decided to focus on the evolution problem: to write the field equations along a $n+1$ foliation of space-time; to write a hamiltonian formulation; to ask the question of the Cauchy problem.

\vspace{\baselineskip}

At first, in the following section, we shall recall the origin of the Lovelock theories and the interest to study the $f($Lovelock$)$ generalisation. Then in the third section we introduce the concepts of evolution problem and Cauchy problem for a covariant theory. We explain in the fourth section how the field equations can be obtained from the action of the theory -- such a proof was not present in the literature before: only the resulting equations were reported in \cite{Bueno16}. In the fifth section, we introduce the formalism usually used to write the $n+1$ decomposition of GR, after which the sixth section contains the calculations of the decomposition for $f($Lovelock$)$. In the seventh section, we summarise the results and apply them to small-degree terms, in order to recover known formulas. The eighth section is the conclusion.

The general decomposed equations did not seem to be known before, and give an original application for the $f(\Rr$, Gauss-Bonnet$)$ case.

\section{Motivations for a \texorpdfstring{$f(\text{Lovelock})$}{f(Lovelock)} theory}

Before introducing $f(\text{Lovelock})$ gravity theories, we have to recall some known results about Lovelock theories. These theories find their origin in two theorems proved in 1971 by D. Lovelock (see \cite{Love71}) after he asked two questions about General Relativity.

\subsection{Lovelock theorems} \label{subsecLovth}
The first question was: could the Einstein equations be different? That is to say, what is the most general 2-tensor that could be involved in gravitational field equations with the same properties as the Einstein tensor, namely: be symmetric, divergent-free, and concomitant of the metric and its first two derivatives $(g,\drond g, \drond^2 g)$?

The symmetry comes from the symmetry of the metric $g$ and from the hypothesis that the space-time affine connection is the Levi-Civita connection of $g$. It can be released if one enables the connection to have a torsion, such as in Einstein-Cartan theory.

The divergence-free condition ensures the conservation of energy.

The limit in the derivatives number comes essentially from the Ostrogradsky's instability: this instability appears for differential equations of order in time higher than two -- see \cite{Lang15} for a nuance on this matter. It is the main reason for which classical dynamics obey to second-order in time motion equations; at the quantum level, it prevents higher-degree theories to be renormalisable and so violates unitarity. Because the theory is supposed to be covariant in space-time, the maximal order of derivatives of the metric is the same as for the time, ie. the second order.

The answer of this first question is: a sum of contractions of several copies of the Riemann tensor. They can be seen as generalisations of the Einstein tensor.

\vspace{\baselineskip}

Then a second question arises naturally: of which lagrangian density can this tensor be the Euler-Lagrange derivative\footnote{Let us remark that the divergence-free condition will automatically be fulfilled by such a tensor. Indeed, it is a classical result (see \cite{Rund66}, or \cite{Wald}) that the field equations deriving from any covariant lagrangian density are divergence-free.}? The answer is not unique, but one convenient candidate is, up to multiplicative constants, a sum of \textbf{scalar} contractions of several copies of the Riemann tensor. The resulting action represents the set of Lovelock theories.

The contraction of $p$ copies of the Riemann tensor is called the $p$-th Lovelock product, and it can be seen as a generalisation of the scalar curvature. In dimension $n+1 = 4$, the Lovelock products' contribution to the action vanishes for $p \geq 2$ and we recover the Einstein equations. The Lovelock theorems hence prove the uniqueness of GR among the theories fulfilling the three required conditions.

\subsection{Geometry issues}
\label{geomissues}

When one releases the dimension, one must be struck by astonishing properties of the Lovelock products. Firstly, they vanish for $2p > n+1$ and thus are in finite number. Secondly, when $n+1$ is even, the last non-vanishing term is nothing but the Gauss-Bonnet-Chern scalar of the space-time. According to Gauss-Bonnet-Chern theorem, it is the integrand of the Euler characteristic of the space-time manifold, which is a topological invariant. Hence it has no contribution in the Lagrangian of the theory. Up to now, the deep connection between the physical assumptions on the Lovelock tensors (symmetric, divergence-free, second-order derivative of the metric) and the geometrical result (Gauss-Bonnet-Chern theorem) is unknown, according to D. Lovelock himself (private communication). This open question may be very interesting to study further.

Because of this coincidence, each Lovelock product promises to have interesting geometric properties, with the particular Gauss-Bonnet-Chern term appearing for the largest non-vanishing $p$, and deserves to be studied for itself.

\subsection{Physics issues}

Apart from this geometrical stake, from the 2000's on the Lovelock products aroused a renewed physical interest and gave birth to a flourishing literature. The string/M-theories describing a high-dimensional space-time recall the original question raised by D. Lovelock and the generalisation of the Einstein tensor in $n > 3$. In these theories, higher-order curvature terms appear naturally (see \cite{Tor08}), such as the 4-dimensional Gauss-Bonnet-Chern term, which is simply called the Gauss-Bonnet term. As we said this term is a topological invariant for $n+1=4$, but not beyond, hence playing a non-trivial role in the field equations. In such a framework, the Lovelock terms could have cosmological implications and could be connected to the dark energy problem: the expansion of the universe might indeed be explained by some Gauss-Bonnet term in the action (see \cite{Nojiri05GBDE}, \cite{Nojiri05MGBDE}, \cite{Nojiri06}, \cite{Cognola06}). This particular case of Lovelock theories containing only the Lovelock products for $p \leq 2$ is usually called Gauss-Bonnet gravity theories.

Even for $n+1=4$, although the Gauss-Bonnet term is the integrand of a topological invariant, this is not the case anymore when it is coupled to a scalar field. The application of a scalar-tensor approach \emph{à la} Brans-Dicke to Gauss-Bonnet theories gives physically different results (see \cite{Tian16}, or \cite{Toloza13} for a model that lets the torsion be different from zero).

At last, the higher-order curvature terms as well as the extra degrees of freedom coming from the scalar fields of a scalar-tensor Lovelock theory give interesting results if seen as a toy-model for the holography context of CFT (see \cite{Bueno16}). This gives the idea of generalising at most the Lovelock theories, using the same procedure as the building of $f(\Rr)$ theories from GR: to put the Lovelock products in an arbitrary function $f$, thus generating the so-called $f(\text{Lovelock})$ theories. Actually, they are a particular case of what could be called scalar-tensor-Lovelock theories, as well as $f(\Rr)$ are a particular case of scalar-tensor theories. As a matter of fact, a Legendre transform makes the $f(\text{Lovelock})$ theories equivalent to coupling each Lovelock product in the action with a scalar field.

\vspace{\baselineskip}

Doing so, we gain on one side in generalisation but we lose on one other: the field equations of $f(\text{Lovelock})$ theories are of fourth-order in the derivatives of $g$.

Indeed, an interesting property of the Lovelock products is that they are a concomitant of $(g,\drond g, \drond^2 g)$. This point is not surprising, insofar as this term has been obtained by two integrations of the Lovelock field equations, which are supposed to be themselves concomitant of the metric and its first two derivatives. However, the peculiarity of the Lovelock products appears when they are compared to other lagrangian densities of the form $L(g,\drond g,\drond^2 g)$. The tensor deriving from such a density is necessarily symmetric and divergence-free\footnote{Symmetric because the Levi-Civita connection is torsion-free; divergence-free as far as $L$ is supposed to be covariant, as explained in subsection \ref{subsecLovth}.}, but it could depend on up to the fourth-order derivative in $g$. Hence the remarkable property of the Lovelock products is that they are in second-order in $g$ \textbf{and so is} their total derivative with respect to $g$.

When one couple them to scalar fields however, one loses this property. One ends up with a general lagrangian density $L(g,\drond g,\drond^2g)$ which gives fourth-order field equations; and we have already explained in subsection \ref{subsecLovth} why second-order field equations are interesting: they imply that the Ostrogradsky's instability is evaded. But is it the only way to escape this instability?

Let us comment on this point.

\vspace{\baselineskip}

The most general second-order field equations, ie. the most general concomitant tensors of $(g, \drond g, \drond^2 g)$ and of scalar fields with their derivatives $(\phi, \drond \phi, \drond^2 \phi)$ are described by the Horndeski theories (see \cite{Horn74}). Obviously, they do not show an Ostrogradsky's instability. They are more complex and contain more coupling terms between the curvature and the scalar field terms than the scalar-tensor-Lovelock theories. Recently, it has been shown that the constraint in Horndeski on the second-order derivative can be released in some extend, hence giving birth to beyond Horndeski theories. Refusing Ostrogradsky's instability actually enables some derivatives higher than 2, under some degeneracy condition on the terms carrying these higher-order derivatives in the Lagrangian. A classification of such Lagrangians is exposed in \cite{Lang15} (see also \cite{Lang17}).

$f(\Rr)$ theories are known to escape the Ostrogradsky's instability. When they are seen as scalar-tensor theories, one can integrate by part their lagrangian density and make the second-order derivatives disappear in the Lagrangian, which is a secure method to evade the instability.

As we explained, Lovelock theories escape the instability as well, because their field equations are at most of second order in time.

What about $f(\text{Lovelock})$ theories? Do they evade Ostrogradsky's instability? The question is much more involved than for the previous two cases and is still open. For the particular $p=2$ case, ie. $f(\Rr$, Gauss-Bonnet$)$, one can find a proof in \cite{Kobayashi11} that the field equations can be written as the field equations of some Horndeski theories. Hence $f(\Rr$, Gauss-Bonnet$)$ theories seem free from the instability (see \cite{Crisos18}).

We won't handle the Ostrogradsky's instability of $f(\text{Lovelock})$ theories in this paper, nor Horndeski theories. We just focus on the $f(\text{Lovelock})$ action, which is the most general action that can be built on the Lovelock products, whose connection with Gauss-Bonnet-Chern theorem promises interesting geometrical stakes. We proceed to the $n+1$ decomposition of these theories which was still to be done.

\section{The Cauchy problem}

Let us recall what the $n+1$ decomposition is about.


\subsection{The Cauchy problem for General Relativity}
In GR, the field equations are covariant: there is neither unique time nor space coordinates in which the equations of motion can be written. The initial value problem, or \textbf{Cauchy problem}, is the question of existence and uniqueness of a solution once given initial data. This problem must then be divided in four parts:
\begin{enumerate}
	\item The choice of a global time coordinate (some time function whose gradient is time-like everywhere) and local space coordinates; this choice gives a space-like foliation of space-time. This is called the $\mathbf{n+1}$ \textbf{decomposition}.
	\item The projection of the field equations in this specific data decomposition. Here the $n+1$-covariance is lost. We obtain two sets of equations: some ones about the geometry of each space-like slice, which are called \textbf{constraint equations}; the others about the evolution of the variables, which are called \textbf{dynamical equations}.
	\item The choice of suitable initial data, ie. some space-like manifold verifying the constraint equations. The resolution of the constraint equations can be formulated as seeking a solution for a non-linear elliptic PDE problem.
	\item The \textbf{well-posedness} of the theory, using the dynamical equations: once the constraint equations are solved and initial data are given, can they be propagated along the time vector field? Locally? Globally? Is this propagation unique? Is it stable under perturbations around a given background? Is the map from the space of initial data to the space of developments smooth? The dynamical equation has to be written in a non-linear hyperbolic system for the Cauchy problem to be well-posed.
\end{enumerate}
It has been shown in 1952 by Y. Choquet-Bruhat in \cite{Cho52} that in the case of vacuum, the Cauchy problem for the Einstein equations is well-posed. It was then proven for different types of matter fields, and for some gravitation theories other than GR. The question arises naturally for Lovelock and $f(\text{Lovelock})$ theories.

\subsection{The Cauchy problem for Lovelock theories}
The first paper studying the hamiltonian formulation of the Lovelock theories is from C. Teitelboim and J. Zanelli (\cite{Tei87}), and does not answer its well-posedness. The seminal work on this topic is from Y. Choquet-Bruhat (\cite{Cho88}, \cite{Cho09-692}), in which the Cauchy theorem is used to establish the well-posedness for small analytical data, given arbitrary lapse and shift. A huge difference with GR appears though, regarding the characteristic hypersurfaces: they are non-null cones, unlike the null cones of GR. Hence gravity can travel slower or faster than light.

In \cite{Deruelle89}, N. Deruelle and L. Fari\~na-Busto showed that cosmological models of Lovelock gravity can present some problematic behaviours, like an universe coming from nowhere or disappearing in a finite time without curvature explosion. These results were gathered in a more general point of view in \cite{Deruelle03}, where the quasi-linearity of the dynamical equations in $\drond^2g$ is showed to be the origin of the pathology.

Nowadays, the current works are on the one hand about the global hyperbolicity of the Lovelock theory. H. S. Reall, N. Tanahashi and B. Way studied the very nature of the characteristic hypersurfaces; they proved in \cite{Reall14} that any Killing horizon is characteristic, and discuss for some cases whether the Lovelock theory is hyperbolic or not. In \cite{Reall15}, they establish that the transport equation along the characteristic hypersurfaces is non-linear, unlike the GR one, and that this non-linearity can lead to shock formation. Finally, the question of the non-linear stability of the Minkowski space-time seems to find a positive answer, although it is not yet proved. On the other hand, S. Willison investigated in \cite{Will14} and \cite{Will15} the quasi-linear reformulation of the Lovelock theory to discuss the local well-posedness of the Cauchy problem with respect to black holes backgrounds.

As for the resolution of the Lovelock constraint equations, quite nothing had be done so far. Insofar as the field equations involve the geometry of the space-time, the constraint equations involve geometrical properties of the space-like data hypersurfaces, such as the \textbf{Yamabe problem} or the {\boldmath $\sigma_k$}\textbf{-curvature} prescription. For the first time we solved in \cite{Lac17-2} the constraint equations in the compact, time-symmetric, conformally flat case for some sets of coefficients.

\subsection{The Cauchy problem for \texorpdfstring{$f(\text{Lovelock})$}{f(Lovelock)} theories}
The decomposition of the Lovelock field equations onto a $n+1$ space-like foliation of space-time is already in the literature (see \cite{Tei87}, \cite{Cho88}, \cite{Cho09-692}); it was made explicit for Gauss-Bonnet gravity in \cite{Tor08}. But the decomposition of $f(\text{Lovelock})$ field equations had not been done up to now: only the field equations were written in \cite{Bueno16} but without the $n+1$ projection. This is the topic of the present paper.

The mathematical resolution of the general $f(\text{Lovelock})$ constraint equations, as well as the well-posedness of the theories, are still completely open questions.

\section{\texorpdfstring{$f(\text{Lovelock})$}{f(Lovelock)} field equations}

In the present section we shall introduce the notations to write the action of a $f($Lovelock$)$ theory, and then we shall show how to derive the field equations.

\subsection{Notations}

We represent the space-time by a lorentzian manifold $(\Vc,g_{\mu\nu})$ of dimension $n+1$, $n \in \Nb$ standing for the spatial dimension. We choose $c=\kappa=1$ for unit and $(-1,+n)$ for the signature of $g_{\mu\nu}$.

We note, borrowing the notations of \cite{Sot10} and \cite{Bar}:
\begin{center}
\begin{tabular}{rp{8cm}}
	\toprule
$D$ & the Levi-Civita connection on $\Vc$, \\
$\Rr_{\mu\nu\rho\sigma}^\Vc$ & the Riemann tensor of $D$, \\
$\square$ & the d'Alembertian on $\Vc$, \\
$\d v_\Vc = \sqrt{-g}\d^{n+1} x$ & the volume element of $\Vc$. \\
	\bottomrule
\end{tabular}
\end{center}
For any tensor $T$ on $\Vc$, $T_{,a}$ stands for $\drond_a T $ and $T_{|a}$ for $D_a T$. The $\Vc$ exponent on the Riemann tensor will be omitted when it is not ambiguous. In its twice covariant and twice contravariant form, using its symmetries, the Riemann tensor can be written \[\Rr_{\gamma \delta}^{\alpha \beta} = \Rr_{\gamma \delta}^{\phantom{\gamma \delta} \alpha \beta} = \Rr_{\phantom{\alpha \beta} \gamma \delta}^{\alpha \beta}\] as well.

We introduce $p_n = \left\lfloor \dfrac{n+1}{2} \right\rfloor$,
\[\delta_{\mu_1 \mu_2\ldots \mu_k}^{\nu_1 \nu_2\ldots \nu_k} := \det
		\begin{pmatrix}
		\delta_{\mu_1}^{\nu_1} & \ldots & \delta_{\mu_1}^{\nu_k} \\
		\vdots						 &				&	\vdots \\
		\delta_{\mu_k}^{\nu_1} & \ldots & \delta_{\mu_k}^{\nu_k}
		\end{pmatrix}\]
the generalised Kronecker symbol,
\begin{center}
\begin{tabular}{ll}
	\toprule
$\Rb_p = \dfrac{1}{2^p}\delta_{\alpha_1 \beta_1 \alpha_2 \beta_2 \cdots \alpha_p \beta_p}^{\gamma_1 \delta_1 \gamma_2 \delta_2 \cdots \gamma_p \delta_p}\Rr_{\gamma_1 \delta_1}^{\alpha_1 \beta_1}\Rr_{\gamma_2 \delta_2}^{\alpha_2 \beta_2} \ldots \Rr_{\gamma_p \delta_p}^{\alpha_p \beta_p}$ & the $p$-th Lovelock product, \\
	\midrule
$\Rb_0 = 1$, \\
$\Rb_1 = \Rr$ & the scalar curvature, \\
$\Rb_2 = \Rr^2 - 4 \Rr_\alpha^\gamma\Rr^\alpha_\gamma + \Rr_{\alpha\beta}^{\gamma\delta}\Rr^{\alpha\beta}_{\gamma\delta}$ & the $n+1=4$ Gauss-Bonnet term, \\
$\Rb_3 =
	\Rr^3
	+2\Rr_{\alpha\beta}^{\gamma\delta}\Rr_{\gamma\delta}^{\varepsilon\eta}\Rr_{\varepsilon\eta}^{\alpha\beta}
	+3\Rr\Rr_{\alpha\beta}^{\gamma\delta}\Rr_{\gamma\delta}^{\alpha\beta}$ \\
	\quad $+8\Rr_{\alpha\beta}^{\gamma\eta}\Rr_{\gamma\delta}^{\varepsilon\beta}\Rr_{\varepsilon\eta}^{\alpha\delta} -12\Rr\Rr_\alpha^\beta\Rr_\beta^\alpha +16\Rr_\alpha^\beta\Rr_\beta^\gamma\Rr_\gamma^\alpha$ \\
	\quad $-24\Rr_\alpha^\beta\Rr_{\gamma\delta}^{\alpha\varepsilon}\Rr_{\beta\varepsilon}^{\gamma\delta}
	+24\Rr_\alpha^\beta\Rr_\gamma^\delta\Rr_{\beta\delta}^{\alpha\gamma}$ & and so on, until \\
$\Rb_p = 0$ for $p > p_n$, & because of the antisymmetries of $\Rr_{\alpha\beta}^{\gamma\delta}$. \\
	\bottomrule
\end{tabular}
\end{center}
We introduce as well
\begin{center}
\begin{tabular}{rl}
	\toprule
$\Rbd_{(p){\mu}}^{\phantom{(p)}{\nu}}$ & $\displaystyle{:= \dfrac{1}{2^p} \delta_{{\mu} \alpha_1 \beta_1 \ldots \alpha_p \beta_p}^{{\nu} \gamma_1 \delta_1 \ldots \gamma_p \delta_p} \Rr_{\gamma_1 \delta_1}^{\alpha_1 \beta_1} \ldots \Rr_{\gamma_p \delta_p}^{\alpha_p \beta_p}}$, \\
	\midrule
$\Rbd_{(0)\mu}^{\phantom{(0)}\nu}$ & $\displaystyle{= \delta_\mu^\nu}$, \\
$\Rbd_{(1)\mu}^{\phantom{(1)}\nu}$ & $\displaystyle{= \Rr \delta_\mu^\nu - 2\Rr_\mu^\nu}$, \\
$\Rbd_{(2)\mu}^{\phantom{(2)}\nu}$ & $\displaystyle{= \left(\Rr^2 - 4 \Rr_\alpha^\gamma\Rr^\alpha_\gamma + \Rr_{\alpha\beta}^{\gamma\delta}\Rr^{\alpha\beta}_{\gamma\delta}\right)\delta_\mu^\nu}$ \\
& \quad $\displaystyle{- 4 \left(\Rr\Rr_\mu^\nu -2 \Rr_\mu^\alpha \Rr_\alpha^\nu - 2\Rr_\alpha^\beta \Rr_{\mu\beta}^{\nu\alpha} + \Rr_{\mu\alpha}^{\beta\gamma}\Rr_{\beta\gamma}^{\nu\alpha}\right)}$, \\
	\bottomrule
\end{tabular}
\end{center}
and
\begin{center}
\begin{tabular}{rl}
	\toprule
$\Rbdd_{(p)\alpha\beta}^{\phantom{(p)}\gamma\delta}$ & $\displaystyle{:= \dfrac{1}{2^p} \delta_{\alpha \beta \alpha_1 \beta_1 \ldots \alpha_p \beta_p}^{\gamma \delta \gamma_1 \delta_1 \ldots \gamma_p \delta_p} \Rr_{\gamma_1 \delta_1}^{\alpha_1 \beta_1} \ldots \Rr_{\gamma_p \delta_p}^{\alpha_p \beta_p}}$, \\
	\midrule
$\Rbdd_{(0)\alpha\beta}^{\phantom{(0)}\gamma\delta}$ & $\displaystyle{= \delta_\alpha^\gamma \delta_\beta^\delta - \delta_\alpha^\delta \delta_\beta^\gamma}$, \\
$\Rbdd_{(1)\alpha\beta}^{\phantom{(1)}\gamma\delta}$ & $\displaystyle{= 2\Rr \left(\delta_\alpha^\gamma \delta_\beta^\delta - \delta_\alpha^\delta \delta_\beta^\gamma\right) - 4\left(\Rr_\alpha^\gamma \delta_\beta^\delta + \Rr_\beta^\delta \delta_\alpha^\gamma - \Rr_\alpha^\delta \delta_\beta^\gamma - \Rr_\beta^\gamma \delta_\alpha^\delta\right) +4\Rr_{\alpha \beta}^{\gamma\delta}}$. \\
	\bottomrule
\end{tabular}
\end{center}

\subsection{Lovelock and Gauss-Bonnet-Chern theorems}
In this paragraph we shall precise the Lovelock theorems and their connection with the Gauss-Bonnet-Chern theorem briefly mentioned in subsection \ref{geomissues}.

The first Lovelock theorem states that the most general form for a symmetric, divergent-free 2-tensor, concomitant of the metric and its first two derivatives $(g,\drond g, \drond^2 g)$ is a linear combination of $\Rbd_{(p)\mu}^{\phantom{(p)}\nu}$:
\[\Ac_\mu^\nu := \sum_{p=0}^{+\infty} \lambda_p \Rbd_{(p)\mu}^{\phantom{(p)}\nu},\]
with $\lambda_p$ real constants.

The second Lovelock theorem shows that this tensor is the Euler-Lagrange derivative of
\[L_\text{Lov} := \sqrt{-g} \sum_{p=0}^{+\infty} \left(-2\lambda_p\right) \Rb_p.\]
We can remark a few things about $L_\text{Lov}$.

Firstly, the sum is finite. Indeed, the symmetries of $\delta$ imply that
\[\text{if } k > n+1, \qquad \delta_{\mu_1 \mu_2\ldots \mu_k}^{\nu_1 \nu_2\ldots \nu_k} = 0,\]
so the terms of the sum vanish for $2p > n+1$.

Secondly, when $n+1$ is even, we observe that
\begin{equation}
\Rbd_{(p_n)\mu}^{\phantom{(p_n)}\nu} = 0 \label{Rpn0}
\end{equation}
because the generalised Kronecker symbol vanishes, of which size is $2p+1$ in $\Rbd_{(p)\mu}^{\phantom{(p)}\nu}$. This means that $\derd{\left(\Rb_{p_n}\sqrt{-g}\right)}{g^{\mu\nu}} = 0$ for all metric $g_{\mu\nu}$, ie.
\[\delta\left(\int_\Vc \Rb_{p_n}\d v_\Vc\right) = 0.\]
In other terms, \textbf{$\Rb_{p_n}$ is a topological invariant}, and does not contribute to the action. This could have been deduced from the Gauss-Bonnet-Chern theorem, which ensures that
\[\int_\Vc \Rb_{p_n}\d v_\Vc = \left(\pi^{p_n}2^{2p_n}p_n!\right)\chi(\Vc),\]
where $\chi(\Vc)$ is the Euler characteristic of $\Vc$. Indeed, $\Rb_{p_n}$ is nothing but the Gauss-Bonnet scalar in dimension $n+1$. However, $(\ref{Rpn0})$ is anything but a proof of this fundamental geometry theorem. The most difficult point in the Gauss-Bonnet-Chern theorem is to prove that $\Rb_{p_n}$ is the integrand of the Euler characteristic, not only that it is a topological invariant.

\subsection{Derivation of the field equations}

As explained in the introduction, we generalise $L_\text{Lov}$ and consider the $f(\text{Lovelock})$ action:
\begin{align}
\label{actionfLov0}
\Sc_{f(\text{Lov})}[g] = \int_\Vc f(\Rb_0,\Rb_1,\ldots,\Rb_{p_n}) \d v_\Vc,
\end{align}
with $f$ an arbitrary function.

The usual procedure for studying $f(\Rr)$ theories is to introduce the Legendre transform of $f$ in order to write the $f(\Rr)$ action $\Sc_{f(\Rr)}$ as the action of a scalar-tensor theory. This method has been known for a long time, see for instance \cite{Wands94} or \cite{Sot10} for a review on $f(\Rr)$ theories. Here, we apply the same procedure and introduce auxiliary fields corresponding to the $\derp{f}{\Rb_p}$. The derivatives in $g_{\mu\nu}$ with a degree higher than 2 will then be carried by those fields.

Under the hypothesis that each $\Rb_p \longmapsto \derp{f}{\Rb_p}$ is bijective (which can be expressed in terms of invertibility of the Hessian $\left(\frac{\drond^2 f}{\drond\Rb_p \drond\Rb_q}\right)_{0 \leq p,q \leq p_n}$, see \cite{Bueno16}), we set:
\begin{align*}
\phi_p &:= \derp{f\left(\Rb_0, \ldots, \Rb_{p_n}\right)}{\Rb_p}, \\
V\left(\phi_0, \ldots, \phi_{p_n}\right) &:= \sum_{p=0}^{p_n}\phi_p \Rb_p(\phi_p) - f\left(\Rb_0(\phi_0), \ldots, \Rb_{p_n}(\phi_{p_n})\right),
\end{align*}
such that
\[\Sc_{f(\text{Lov})}[g] = \int_\Vc \left[\sum_{p=0}^{p_n}\phi_p(\Rb_p)\Rb_p - V\left(\phi_0(\Rb_0), \ldots, \phi_{p_n}(\Rr_{p_n})\right) \right] \d v_\Vc.\]
In order to carry out the ADM decomposition of the field equations of this action, we shall derive them while considering the $\phi_p$'s as independent fields. These new degrees of freedom embed the space-time $\Vc$ into a larger phase space, and the classical trajectories correspond to the $p_n+1$ equations: $\phi_p = \derp{f}{\Rr_p}$. Hence the action becomes
\begin{equation}
\label{actionfLovSTT0}
\Sc_{f(\text{Lov})}[g,\phi_0, \ldots, \phi_{p_n}] = \int_\Vc \left[\sum_{p=0}^{p_n}\phi_p\Rb_p - V\left(\phi_0, \ldots, \phi_{p_n}\right) \right] \d v_\Vc.
\end{equation}

The field equations of this action can be found in \cite{Bueno16}, but here we explain how to obtain them. We can use a theorem proved by H. Rund in \cite{Rund66}: the total derivative with respect to $g_{\mu\nu}$ of a lagrangian density $L[g]=L(g,\drond g,\drond^2 g)$, which is given by
\begin{align*}
E^{\mu\nu}
	&:= \derd{L}{g_{\mu\nu}} \\
	&= \derp{L}{g_{\mu\nu}} - \drond_\rho \derp{L}{g_{\mu\nu,\rho}} + \drond_\sigma \drond_\rho \derp{L}{g_{\mu\nu,\rho\sigma}},
\end{align*}
can be expressed by
\[E^{\mu\nu} = \Lambda^{\mu\nu,\rho\sigma}_{\phantom{\mu\nu,\rho\sigma}|\rho\sigma} + \dfrac{1}{2}L g^{\mu\nu} - \dfrac{2}{3}\Rr_{\sigma\phantom{\nu}\tau\rho}^{\phantom{\sigma}\nu}\Lambda^{\rho\sigma,\mu\tau}\]
up to total derivatives, where
\[\Lambda^{\alpha\beta,\gamma\delta} := \derp{L}{g_{\alpha\beta,\gamma\delta}}.\]
So if we set
\begin{center}
\begin{tabular}{rlcrl}
	\toprule
$\Lc_{p}$ & $\displaystyle{:= \phi_p \Rb_{p}\sqrt{-g}}$, &&
$L_{p}$ & $\displaystyle{:= \Rb_{p}\sqrt{-g}}$, \\
	\midrule
$\Ec_{(p)}^{\phantom{(p)}\mu\nu}$ & $\displaystyle{:= \derd{\Lc_{p}}{g_{\mu\nu}}}$, &&
$E_{(p)}^{\phantom{(p)}\mu\nu}$ & $\displaystyle{:= \derd{L_{p}}{g_{\mu\nu}}}$, \\
	\midrule
&&& $\Lambda_{(p)}^{\alpha\beta,\gamma\delta}$ & $\displaystyle{:= \derp{L_p}{g_{\alpha\beta,\gamma\delta}}}$, \\
	\bottomrule
\end{tabular}
\end{center}
we get
\[\Ec_{(p)}^{\phantom{(p)}\mu\nu} = \phi_p E_{(p)}^{\phantom{(p)}\mu\nu} + \Lambda_{(p)}^{\mu\nu,\rho\sigma}\phi_{p|\rho\sigma} + \Lambda^{\mu\nu,\rho\sigma}_{(p)\phantom{\rho\sigma}|\sigma} \phi_{p|\rho} + \Lambda^{\mu\nu,\rho\sigma}_{(p)\phantom{\rho\sigma}|\rho} \phi_{p|\sigma}.\]
Yet, it is showed in \cite{Love71} using Bianchi identities, that
\[\Lambda^{\mu\nu,\rho\sigma}_{(p)\phantom{\rho\sigma}|\sigma} = \Lambda^{\mu\nu,\rho\sigma}_{(p)\phantom{\rho\sigma}|\rho} = 0.\]
Furthermore, the second Lovelock theorem (see \cite{Love71}) states that
\[E_{(p)\mu}^{\phantom{(p)}\nu} = \dfrac{1}{2} \Rbd_{(p)\mu}^{\phantom{(p)}\nu}\sqrt{-g}.\]
Now, taking the value of $\Lambda_{(p)}^{\mu\nu,\rho\sigma}$ given in \cite{Love71}, we can show that
\begin{align*}
\Lambda_{(p)}^{\mu\nu,\rho\sigma}
	&= \dfrac{1}{2}p\sqrt{-g}\left(P_{(p)}^{\mu\rho\sigma\nu} + P_{(p)}^{\mu\sigma\rho\nu} + P_{(p)}^{\nu\rho\sigma\mu} + P_{(p)}^{\nu\sigma\rho\mu}\right) \\
	&= p\sqrt{-g}\left(P_{(p)}^{\mu\rho\sigma\nu} + P_{(p)}^{\mu\sigma\rho\nu}\right),
\end{align*}
where
\begin{align}
P_{(p)}^{\alpha\beta\gamma\delta}
  &:= \dfrac{1}{2^p} \delta_{\alpha_1 \beta_1 \ldots \alpha_{p-1} \beta_{p-1} \alpha_p \beta_p}^{\gamma_1 \delta_1 \ldots \gamma_{p-1} \delta_{p-1} \ \alpha \ \beta} \Rr_{\gamma_1 \delta_1}^{\alpha_1 \beta_1} \ldots \Rr_{\gamma_{p-1} \delta_{p-1}}^{\alpha_{p-1} \beta_{p-1}}g^{\alpha_p \gamma} g^{\beta_p \delta} \nonumber \\
  &= \dfrac{1}{2} \Rbdd_{(p-1)\mu\nu}^{\phantom{(p-1)}\alpha\beta} g^{\mu \gamma} g^{\nu \delta} \label{defP}
\end{align}
is such that
\[\Rb_p = P_{(p)}^{\alpha\beta\gamma\delta}\Rr_{\alpha\beta\gamma\delta}.\]
$P_{(p)}^{\alpha\beta\gamma\delta}$, by homogeneity (see \cite{Pad13}), can be defined as well by
\[P_{(p)}^{\alpha\beta\gamma\delta} = \dfrac{1}{p}\derp{\Rb_p}{\Rr_{\alpha\beta\gamma\delta}}.\]
Hence,
\begin{equation}
\Ec_{(p)}^{\phantom{(p)}\mu\nu} = \dfrac{\phi_p}{2} \Rbd_{(p)}^{\mu\nu}\sqrt{-g} +p\left(P_{(p)}^{\mu\rho\sigma\nu} + P_{(p)}^{\mu\sigma\rho\nu}\right)\phi_{p|\rho\sigma}\sqrt{-g}.
\end{equation}

Now we are ready to derive the field equations. With the addition of a standard matter action $\Sc_\text{mat}$ depending on matter and energy fields represented by $\Psi$, the action $\Sc_{f(\text{Lov})}$ becomes
\begin{align}
\Sc[g,\phi_0, \ldots, \phi_{p_n},\Psi]
	&:= \Sc_{f(\text{Lov})}[g,\phi_0, \ldots, \phi_{p_n}] + \Sc_\text{mat}[g,\Psi] \label{actionfLov} \\
	&= \int_\Vc \left[\sum_{p=0}^{p_n} \Lc_p - V\sqrt{-g} + S_\text{mat}\right] \d^{n+1} x. \nonumber
\end{align}
If we consider a small variation of the metric, $\delta g_{\mu\nu}$, we get:
\begin{align*}
\delta\Sc[g,\phi_0, &\ldots, \phi_{p_n},\Psi]
	= \delta\int_\Vc \left[\sum_{p=0}^{p_n} \Lc_p - V\sqrt{-g} + S_\text{mat} \right] \d^{n+1} x \\
	&= \int_\Vc \left[\sum_{p=0}^{p_n} \left(\dfrac{\phi_p}{2}\Rbd_{(p)}^{\mu\nu} +p\left(P_{(p)}^{\mu\rho\sigma\nu} + P_{(p)}^{\mu\sigma\rho\nu}\right)\phi_{p|\rho\sigma}\right) - \dfrac{1}{2}V g^{\mu\nu} + \dfrac{1}{2} T^{\mu\nu} \right]\sqrt{-g}\delta g_{\mu\nu} \d^{n+1} x,
\end{align*}
where
\begin{equation}
T_{\mu\nu} := -\dfrac{2}{\sqrt{-g}} \derd{S_{\text{mat}}}{g^{\mu\nu}} \label{eqTmunu}
\end{equation}
is the stresss-energy tensor associated to $S_\text{mat}[g,\Psi]$. So $g_{\mu\nu}$ is a critical point of $\Sc$ iff
\begin{equation}
\label{eqchfLL}
\Ac^{\mu\nu} = V g^{\mu\nu} - T^{\mu\nu},
\end{equation}
where
\begin{align}
\Ac^{\mu\nu} &:= \sum_{p=0}^{p_n} \Ac_{(p)}^{\mu\nu}, \\
\Ac_{(p)}^{\mu\nu} &:= \phi_p\Rbd_{(p)}^{\mu\nu} +2p\left(P_{(p)}^{\mu\rho\sigma\nu} + P_{(p)}^{\mu\sigma\rho\nu}\right)\phi_{p|\rho\sigma}.
\end{align}

In order to complete the set of equations, we have to compute the Euler-Lagrange equations with respect to the $\phi_p$'s:
\begin{equation}
\derp{}{\phi_p}\left[\sum_{p=0}^{p_n} \Lc_p - V\sqrt{-g} + S_\text{mat} \right] = D_\alpha \left(\derp{}{D_\alpha \phi_p}\left[\sum_{p=0}^{p_n} \Lc_p - V\sqrt{-g} + S_\text{mat} \right]\right),
\end{equation}
ie.
\begin{equation}
\label{eqchfLLphi}
\Rb_p = \derp{}{\phi_p}V(\phi_0,\phi_1,\ldots,\phi_{p_n}).
\end{equation}
But only the first field equations \eqref{eqchfLL} are concerned by the hamiltonian formulation of $f(\text{Lovelock})$.

\section{\texorpdfstring{$n+1$}{n+1} formalism for General Relativity}

Now that we have the field equations, we shall project them along a $n+1$ decomposition of space-time.

Let us suppose that $\Vc$ is foliated by the level hypersurfaces of a time function $t$, whose gradient is time-like future-oriented everywhere. Let $(\Mc,\gamma_{ij})$ be a level hypersurface of $t$. $\Mc$ is a space-like submanifold of $(\Vc,g_{\mu\nu})$, endowed with its riemannian metric $\gamma_{ij}$ induced by $g_{\mu\nu}$.

We note, borrowing once again the notations of \cite{Sot10} and \cite{Bar}:
\begin{center}
\begin{tabular}{rp{8cm}}
	\toprule
$(t=:x^0,x^i)$ & the local coordinates on $\Vc$, \\
	\midrule
$\nabla$ & the Levi-Civita connection on $\Mc$, \\
$\Rr_{ijkl}^\Mc$ & the Riemann tensor of $\nabla$, \\
$\Delta$ & the Laplacian on $\Mc$, \\
$\d v_\Mc = \sqrt{\gamma}\d^{n} x$ & the volume element of $\Mc$. \\
	\bottomrule
\end{tabular}
\end{center}
Latin indices run from 1 to $n$, while greek ones run from 0 to $n$. The $\Mc$ exponent on the Riemann tensor will be omitted when it is not ambiguous.

We write
\begin{center}
\begin{tabular}{rp{8cm}}
	\toprule
$u^\mu$ & the future-oriented timelike unit normal to $\Mc$, \\
	\midrule
$u^\mu =: \dfrac{1}{N}\left(1,-X^i\right)$ & the coordinates of $u^\mu$ in the basis $(\drond_t, \drond_i)$, \\
	\midrule
$N$ & the lapse, \\
	\midrule
$X^i$ & the shift. \\
	\midrule
$\varepsilon := u^\mu u_\mu = \pm 1$ & Here, $u^\mu$ is timelike, so  $\varepsilon=-1$.

If $u^\mu$ were space-like, we would have $\varepsilon = +1$. \\
	\midrule
$u_\mu = (\varepsilon N,0)$ & the 1-form associated to $u^\mu$ through $g_{\mu\nu}$. \\
	\bottomrule
\end{tabular}
\end{center}

The point to keep $\varepsilon$ instead of $-1$ is that it keeps visible the geometrical origin of the several $-1$ which will appear in the next equations. We borrow this notation to \cite{Tor08}.

\begin{center}
\begin{tabular}{rp{8cm}}
	\toprule
$\gamma_\mu^\nu := g_\mu^\nu - \varepsilon u_\mu u^\nu$ & The projector from $T\Vc$ onto $T\Mc$.

In particular,
\[\gamma_\mu^\nu u^\mu = 0, \qquad \gamma_\mu^\nu \gamma^\mu_\rho = \gamma^\nu_\rho, \qquad \gamma_i^\nu = g_i^\nu = \delta_i^\nu.\]
$\gamma_\mu^\nu$ can be used to link $D$ with $\nabla$: if $T_{\alpha_1\ldots}^{\phantom{\alpha_1\ldots}\beta_1\ldots}$ is a tensor, we have
\[\nabla_\mu T_{\alpha_1\ldots}^{\phantom{\alpha_1\ldots}\beta_1\ldots} = \gamma_\mu^\nu \gamma_{\alpha_1}^{\mu_1} \ldots \gamma_{\nu_1}^{\beta_1} \ldots D_\nu T_{\mu_1\ldots}^{\phantom{\mu_1\ldots}\nu_1\ldots}.\] \\
	\midrule
$K_{\mu\nu} := \nabla_\mu u_\nu = \gamma_\mu^\alpha \gamma_\nu^\beta D_\alpha u_\beta$ & The second fundamental form, or extrinsic curvature, of $\Mc$.

In particular,
\[K_{\mu\nu} = \dfrac{1}{2}\Lc_u\gamma_{\mu\nu}.\]
\\
	\bottomrule
\end{tabular}
\end{center}


Now, we can write the $n+1$ decomposition, or Arnowitt-Deser-Misner (ADM) formulation of $f(\text{Lovelock})$. It consists in decoupling the field equations \eqref{eqchfLL} on $\Mc$ and $u^\mu$. We introduce:
\begin{center}
\begin{tabular}{llp{8cm}}
	\toprule
$E$ & $:= T_{\alpha\beta}u^\alpha u^\beta$ & the energy density. \\
	\midrule
$J_\mu$ & $:= \varepsilon\gamma_\mu^\alpha T_{\alpha\beta}u^\beta$ & the momentum density. \\
	\midrule
$S_{\mu\nu}$ & $:= \gamma_\mu^\alpha\gamma_\nu^\beta T_{\alpha\beta}$ & the stress tensor density. \\
	\bottomrule
\end{tabular}
\end{center}
Such that:
\[T_{\mu\nu} = S_{\mu\nu} + J_\mu u_\nu + J_\nu u_\mu + E u_\mu u_\nu.\]

The same decomposition can be done on $\Rr_{\mu\nu\rho\sigma}^\Vc$, see \cite{Tor08}. Indeed, we have:
\begin{center}
\begin{tabular}{lrl}
	\toprule
The Gauss equation:
	& $\gamma_\alpha^\mu \gamma_\beta^\nu \gamma_\gamma^\rho \gamma_\delta^\sigma \Rr_{\mu\nu\rho\sigma}^\Vc$
	& $= \Rr_{\alpha\beta\gamma\delta}^\Mc - \varepsilon\left(K_{\alpha\gamma}K_{\beta\delta} - K_{\alpha\delta}K_{\beta\gamma}\right)$ \\
	&& $=: \Mr_{\alpha\beta\gamma\delta}$. \\ 
	\midrule
The Codazzi identity:
	& $\varepsilon\gamma_\alpha^\mu \gamma_\beta^\nu \gamma_\gamma^\rho \Rr_{\mu\nu\rho\sigma}^\Vc u^\sigma$
	& $= \varepsilon\left(\nabla_\alpha K_{\beta\gamma} - \nabla_\beta K_{\alpha\gamma}\right)$ \\
	&& $=: \Nr_{\alpha\beta\gamma}$. \\ 
	\midrule
The Mainardi equation:
	& $\gamma_\alpha^\mu \gamma_\gamma^\rho \Rr_{\mu\nu\rho\sigma}^\Vc u^\nu u^\sigma$
	& $= -\Lc_u K_{\alpha\gamma} + K_{\alpha \tau}K^\tau_\gamma - \varepsilon \dfrac{\nabla_\alpha \nabla_\gamma N}{N}$ \\
	&& $=: \Or_{\alpha\gamma}$. \\ 
	\bottomrule
\end{tabular}
\end{center}
$\Nr_{\alpha\beta\gamma}$ is antisymmetric in $\alpha \leftrightarrow \beta$, while $\Or_{\alpha\gamma}$ is symmetric in $\alpha \leftrightarrow \gamma$.

We write:
\begin{align*}
\Mr_{\alpha \beta}^{\gamma \delta} &:= \Mr_{\alpha \beta}^{\phantom{\alpha \beta}\gamma \delta} = \Mr_{\phantom{\gamma \delta}\alpha \beta}^{\gamma \delta},
&\Nr_{\alpha\beta}^\gamma &:= \Nr_{\alpha\beta}^{\phantom{\alpha\beta}\gamma}, \\
\Or_\alpha^\beta &:= \Or_\alpha^{\phantom{\alpha}\beta} = \Or_{\phantom{\beta}\alpha}^\beta,
&\Nr^{\alpha\beta}_\gamma &:= \Nr^{\alpha\beta}_{\phantom{\alpha\beta}\gamma}.
\end{align*}

If then we set:
\begin{align*}
\Nrb_{\mu\nu\rho\sigma} &:= \Nr_{\mu\nu\rho}u_\sigma + \Nr_{\nu\mu\sigma}u_\rho + \Nr_{\rho\sigma\mu}u_\nu + \Nr_{\sigma\rho\nu}u_\mu, \\
\Orb_{\mu\nu\rho\sigma} &:= \Or_{\mu\rho}u_\nu u_\sigma - \Or_{\mu\sigma}u_\nu u_\rho - \Or_{\nu\rho}u_\mu u_\sigma + \Or_{\nu\sigma}u_\mu u_\rho,
\end{align*}
we obtain:
\begin{equation}
\Rr^\Vc_{\mu\nu\rho\sigma} = \Mr_{\mu\nu\rho\sigma} + \Nrb_{\mu\nu\rho\sigma} + \Orb_{\mu\nu\rho\sigma},
\label{decompoRadm}
\end{equation}
where $\Mr_{\alpha\beta\gamma\delta}$, $\Nrb_{\alpha\beta\gamma\delta}$ and $\Orb_{\alpha\beta\gamma\delta}$ have the symmetries of $\Rr_{\alpha\beta\gamma\delta}$. Let us compute the contractions of these tensors with respect to $g_{\mu\nu}$:
\begin{align*}
\Mr_{\mu\rho} &:= g^{\nu\sigma}\Mr_{\mu\nu\rho\sigma} = \gamma^{\nu\sigma}\Mr_{\mu\nu\rho\sigma} = \Rr^\Mc_{\mu\rho} - \varepsilon\left(K K_{\mu\rho} - K_{\mu\tau}K^\tau_\rho\right), \\
\Mr &:= g^{\mu\rho}\Mr_{\mu\rho} = \Rr^\Mc - \varepsilon\left(K^2 - K_{\alpha\beta}K^{\alpha\beta}\right), \\
\Nr_\nu &:= g^{\mu\rho}\Nr_{\mu\nu\rho} = \varepsilon\left(D_\tau K_\nu^\tau - D_\nu K\right), \\
\Nrb_{\mu\rho} &:= g^{\nu\sigma}\Nrb_{\mu\nu\rho\sigma} = \Nr_\mu u_\rho + \Nr_\rho u_\mu, \\
\Nrb &:= g^{\mu\rho}\Nrb_{\mu\rho} = 0, \\
\Or &:= g^{\mu\rho}\Or_{\mu\rho} = -\Lc_u K - K_\mu^\nu K_\nu^\mu - \varepsilon \frac{\Delta N}{N}, \\
\Orb_{\mu\rho} &:= g^{\nu\sigma}\Orb_{\mu\nu\rho\sigma} = \varepsilon\Or_{\mu\rho} + \Or u_\mu u_\rho, \\
\Orb &:= g^{\mu\rho}\Orb_{\mu\rho} = 2\varepsilon\Or.
\end{align*}
We precise as well the further relations:
\begin{align*}
u^\mu\Nrb_{\mu\nu\rho\sigma} &= \varepsilon \Nr_{\sigma\rho\nu}, \\
u^\mu\Nrb_{\mu\rho} &= \varepsilon \Nr_{\rho}, \\
u^\mu\Orb_{\mu\nu\rho\sigma} &= -\varepsilon \Or_{\nu\rho} u_\sigma + \varepsilon \Or_{\nu\sigma} u_\rho, \\
u^\mu\Orb_{\mu\rho} &= \varepsilon \Or u_{\rho}.
\end{align*}

And finally, we set:
\begin{center}
\begin{tabular}{rp{8cm}}
	\toprule
$\Pi_p := \Lc_u\phi_p = u(\phi_p) = u^\mu D_\mu\phi_p$ & The momentum of $\phi_p$, ie. the Lie derivative along $u^\mu$. \\
	\bottomrule
\end{tabular}
\end{center}
Then, we have
\begin{equation*}
D_\alpha D_\beta \phi_p =
\nabla_\alpha \nabla_\beta \phi_p + \varepsilon \Pi_p K_{\alpha\beta} +
\varepsilon u_\alpha u^\mu D_\mu D_\beta \phi_p +
\varepsilon u_\beta u^\nu D_\alpha D_\nu \phi_p -
u_\alpha u_\beta u^\mu u^\nu D_\mu D_\nu \phi_p,
\end{equation*}
hence
\begin{align}
\gamma_\mu^\alpha \gamma_\beta^\nu D_\alpha D^\beta \phi_p
	= \gamma_\mu^\alpha \gamma_\beta^\nu \phi_{p|\alpha}^{\phantom{p}|\beta}
  &= \nabla_\mu \nabla^\nu \phi_p + \varepsilon \Pi_p K_{\mu}^{\nu}, \label{ggDDphi} \\
u^\alpha\gamma_\beta^\nu D_\alpha D^\beta \phi_p
	= u^\alpha\gamma_\beta^\nu \phi_{p|\alpha}^{\phantom{p}|\beta}
  &= \nabla^\nu \Pi_p - K^{\nu\mu}\nabla_\mu \phi_p, \label{ugDDphi} \\
u^\alpha u_\beta D_\alpha D^\beta \phi_p
	= u^\alpha u_\beta \phi_{p|\alpha}^{\phantom{p}|\beta}
  &= \varepsilon\left(\square \phi_p - \Delta \phi_p\right) - \Pi_p K. \label{uuDDphi} \\
  &= \drond_{tt}^2 \phi_p - \Pi_p K. \nonumber \\
\end{align}

\section{\texorpdfstring{$f(\text{Lovelock})$}{f(Lovelock)} decomposition: notations and outline of the calculation} \label{sectionoutline}
\subsection{Notations}

The aim of this paper is to compute the three different projections of \eqref{eqchfLL} on $\Mc \oplus u$:
\begin{align}
\Ac^{\mu\nu}u_\mu u_\nu &= \varepsilon V - E, &\text{(hamiltonian constraint)} \label{eqcontrainte1fLL} \\
\Ac^{\mu\nu}\gamma_{i\mu} u_\nu &= -\varepsilon J_i, &\text{(momentum constraint)} \label{eqcontrainte2fLL} \\
\Ac^{\mu\nu}\gamma_{i\mu} \gamma_{j\nu} &= V \gamma_{ij} - S_{ij}. &\text{(dynamical equations)} \label{eqdynfLL}
\end{align}

The first two lines are called constraint equations. They are necessary conditions for a given hypersurface $(\Mc,\gamma_{ij})$, its extrinsic curvature $K_{ij}$, and scalar fields $\phi_{p}$'s to be the restriction on a slice of a space-time $(\Vc,g_{\mu\nu})$ with scalar fields $\phi_p$'s verifying the field equations $(\ref{eqchfLL}, \ref{eqchfLLphi})$. They only involve quantities restricted to $\Mc$, namely $(\gamma_{ij},K_{ij},\phi_p,\Pi_p)$. In particular, they do not depend explicitly on the lapse $N$ nor the shift $X^i$. The ``outgoing'' motion, ie. the dependence in time, is encoded in $K_{\mu\nu} = \frac{1}{2}\Lc_u \gamma_{\mu\nu}$ and $\Pi_p = \Lc_u \phi_p$. Hence in $(\gamma_{ij},K_{ij},\phi_p,\Pi_p)$ we recognize a canonical hamiltonian structure $(q,p)$ lying on $\Mc$.

The third line, the dynamical equations, are more involved: they describe the evolution of $(\gamma_{ij},K_{ij},\phi_p,\Pi_p)$ along $u^\mu$, and therefore along $\drond_t$. They have to contain explicit mentions of the lapse $N$. They can either be seen as first-order in time equations on $(\gamma_{ij},K_{ij},\phi_p,\Pi_p)$, or second-order in time equations on $(\gamma_{ij},\phi_p)$ through $\square \phi_p$ for $\phi_p$ and $\Lc_u K_{\mu\nu}$ for $\gamma_{\mu\nu}$.

We intend to highlight all these dependences. So we have to rewrite $(\ref{eqcontrainte1fLL},\ref{eqcontrainte2fLL})$ without explicit mention of time: all the $u^\mu$'s have to disappear. Likewise, the dependences in time of \eqref{eqdynfLL} have to be confined to $\square \phi_p$ (or $\drond_{tt}^2 \phi_p$) and $\Lc_u K_{\mu\nu}$. For this purpose, we have to introduce
\[\begin{array}{ll}
\toprule
\Mbdd_{(p)\alpha\beta}^{\phantom{(p)}\gamma\delta} &:= \dfrac{1}{2^p} \delta_{\alpha_1 \beta_1 \ldots \alpha_p \beta_p \alpha \beta}^{\gamma_1 \delta_1 \ldots . \gamma_p \delta_p \gamma \, \delta} \Mr_{\gamma_1 \delta_1}^{\alpha_1 \beta_1} \ldots \Mr_{\gamma_p \delta_p}^{\alpha_p \beta_p}, \\
\Mbd_{(p)\alpha}^{\phantom{(p)}\gamma} &:= \dfrac{1}{2^p} \delta_{\alpha_1 \beta_1 \ldots \alpha_p \beta_p \alpha}^{\gamma_1 \delta_1 \ldots . \gamma_p \delta_p \gamma} \Mr_{\gamma_1 \delta_1}^{\alpha_1 \beta_1} \ldots \Mr_{\gamma_p \delta_p}^{\alpha_p \beta_p}, \\
\Mb_{(p)} &:=
\dfrac{1}{2^p} \delta_{\alpha_1 \beta_1 \ldots \alpha_p \beta_p}^{\gamma_1 \delta_1 \ldots \gamma_p \delta_p} \Mr_{\gamma_1 \delta_1}^{\alpha_1 \beta_1} \ldots \Mr_{\gamma_p \delta_p}^{\alpha_p \beta_p}, \\
\Mr_{\gamma\delta}^{\alpha\beta} &=
\Rr_{\phantom{\Mc}\gamma\delta}^{\Mc\alpha\beta} - \varepsilon\left(K_\gamma^\alpha K_\delta^\beta - K_\delta^\alpha K_\gamma^\beta \right), \\
\midrule
\Nbdd_{(p)\mu\alpha\beta}^{\phantom{(p)\mu}\gamma\delta} &:=
\dfrac{1}{2^p}\delta_{\alpha_1 \beta_1 \ldots \alpha_{p-1} \beta_{p-1} \alpha_p \mu \alpha \beta}^{\gamma_1 \delta_1 \ldots . \gamma_{p-1} \delta_{p-1} \gamma_p \delta_p \gamma \delta} \Mr_{\gamma_1 \delta_1}^{\alpha_1 \beta_1} \ldots \Mr_{\gamma_{p-1} \delta_{p-1}}^{\alpha_{p-1} \beta_{p-1}} \Nr_{\gamma_p \delta_p}^{\alpha_p}, \\
\Nbd_{(p)\mu\alpha}^{\phantom{(p)\mu}\gamma} &:=
\dfrac{1}{2^p}\delta_{\alpha_1 \beta_1 \ldots \alpha_{p-1} \beta_{p-1} \alpha_p \mu \alpha}^{\gamma_1 \delta_1 \ldots . \gamma_{p-1} \delta_{p-1} \gamma_p \delta_p \gamma} \Mr_{\gamma_1 \delta_1}^{\alpha_1 \beta_1} \ldots \Mr_{\gamma_{p-1} \delta_{p-1}}^{\alpha_{p-1} \beta_{p-1}} \Nr_{\gamma_p \delta_p}^{\alpha_p}, \\
\Nb_{(p)\mu} &:=
\dfrac{1}{2^p}\delta_{\alpha_1 \beta_1 \ldots \alpha_{p-1} \beta_{p-1} \alpha_p \mu}^{\gamma_1 \delta_1 \ldots \gamma_{p-1} \delta_{p-1} \gamma_p \delta_p} \Mr_{\gamma_1 \delta_1}^{\alpha_1 \beta_1} \ldots \Mr_{\gamma_{p-1} \delta_{p-1}}^{\alpha_{p-1} \beta_{p-1}} \Nr_{\gamma_p \delta_p}^{\alpha_p}, \\
\Nr_{\gamma \delta}^\alpha &=
\varepsilon\left(\nabla_\gamma K_\delta^\alpha - \nabla_\delta K_\gamma^\alpha\right), \\
\midrule
\Nbdd_{(p)\phantom{\nu}\alpha\beta}^{\phantom{(p)}\nu\gamma\delta} &:=
\dfrac{1}{2^p}\delta_{\alpha_1 \beta_1 \ldots \alpha_{p-1} \beta_{p-1} \alpha_p \beta_p \alpha \beta}^{\gamma_1 \delta_1 \ldots . \gamma_{p-1} \delta_{p-1} \gamma_p \ \nu \, \gamma \, \delta} \Mr_{\gamma_1 \delta_1}^{\alpha_1 \beta_1} \ldots \Mr_{\gamma_{p-1} \delta_{p-1}}^{\alpha_{p-1} \beta_{p-1}} \Nr_{\gamma_p}^{\alpha_p \beta_p}, \\
\Nbd_{(p)\phantom{\nu}\alpha}^{\phantom{(p)}\nu\gamma} &:=
\dfrac{1}{2^p}\delta_{\alpha_1 \beta_1 \ldots \alpha_{p-1} \beta_{p-1} \alpha_p \beta_p \alpha}^{\gamma_1 \delta_1 \ldots . \gamma_{p-1} \delta_{p-1} \gamma_p \ \nu \, \gamma} \Mr_{\gamma_1 \delta_1}^{\alpha_1 \beta_1} \ldots \Mr_{\gamma_{p-1} \delta_{p-1}}^{\alpha_{p-1} \beta_{p-1}} \Nr_{\gamma_p}^{\alpha_p \beta_p}, \\
\Nb_{(p)\phantom{\nu}}^{\phantom{(p)}\nu} &:=
\dfrac{1}{2^p}\delta_{\alpha_1 \beta_1 \ldots \alpha_{p-1} \beta_{p-1} \alpha_p \beta_p}^{\gamma_1 \delta_1 \ldots . \gamma_{p-1} \delta_{p-1} \gamma_p \ \nu} \Mr_{\gamma_1 \delta_1}^{\alpha_1 \beta_1} \ldots \Mr_{\gamma_{p-1} \delta_{p-1}}^{\alpha_{p-1} \beta_{p-1}} \Nr_{\gamma_p}^{\alpha_p \beta_p}, \\
\Nr_{\gamma}^{\alpha \beta} &=
\varepsilon\left(\nabla^\alpha K_\gamma^\beta - \nabla^\beta K_\gamma^\alpha\right), \\
\midrule
\NNbdd_{(p)\alpha\beta}^{\phantom{(p)}\gamma\delta} &:=
\dfrac{1}{2^p}\delta_{\alpha_1 \beta_1 \ldots \alpha_{p-1} \beta_{p-1} \alpha_{p} \alpha_{p+1} \beta_{p+1} \alpha \beta}^{\gamma_1 \delta_1 \ldots . \gamma_{p-1} \delta_{p-1} \gamma_{p} \, \delta_{p\phantom{+1}} \gamma_{p+1} \gamma \, \delta} \Mr_{\gamma_1 \delta_1}^{\alpha_1 \beta_1} \ldots \Mr_{\gamma_{p-1} \delta_{p-1}}^{\alpha_{p-1} \beta_{p-1}} \Nr_{\gamma_{p} \delta_{p}}^{\alpha_{p}} \Nr_{\gamma_{p+1}}^{\alpha_{p+1} \beta_{p+1}}, \\
\NNbd_{(p)\alpha}^{\phantom{(p)}\gamma} &:=
\dfrac{1}{2^p}\delta_{\alpha_1 \beta_1 \ldots \alpha_{p-1} \beta_{p-1} \alpha_{p} \alpha_{p+1} \beta_{p+1} \alpha}^{\gamma_1 \delta_1 \ldots . \gamma_{p-1} \delta_{p-1} \gamma_{p} \, \delta_{p\phantom{+1}} \gamma_{p+1} \gamma} \Mr_{\gamma_1 \delta_1}^{\alpha_1 \beta_1} \ldots \Mr_{\gamma_{p-1} \delta_{p-1}}^{\alpha_{p-1} \beta_{p-1}} \Nr_{\gamma_{p} \delta_{p}}^{\alpha_{p}} \Nr_{\gamma_{p+1}}^{\alpha_{p+1} \beta_{p+1}}, \\
\NNb_{(p)} &:=
\dfrac{1}{2^p}\delta_{\alpha_1 \beta_1 \ldots \alpha_{p-1} \beta_{p-1} \alpha_{p} \alpha_{p+1} \beta_{p+1}}^{\gamma_1 \delta_1 \ldots . \gamma_{p-1} \delta_{p-1} \gamma_{p} \, \delta_{p\phantom{+1}} \gamma_{p+1}} \Mr_{\gamma_1 \delta_1}^{\alpha_1 \beta_1} \ldots \Mr_{\gamma_{p-1} \delta_{p-1}}^{\alpha_{p-1} \beta_{p-1}} \Nr_{\gamma_{p} \delta_{p}}^{\alpha_{p}} \Nr_{\gamma_{p+1}}^{\alpha_{p+1} \beta_{p+1}}, \\
\midrule
\Obdd_{(p)\alpha\beta}^{\phantom{(p)}\gamma\delta} &:=
\dfrac{1}{2^p}\delta_{\alpha_1 \beta_1 \ldots \alpha_{p} \beta_{p} \alpha_{p+1} \alpha \beta}^{\gamma_1 \delta_1 \ldots . \gamma_{p} \delta_{p} \gamma_{p+1} \gamma \delta} \Mr_{\gamma_1 \delta_1}^{\alpha_1 \beta_1} \ldots \Mr_{\gamma_{p} \delta_{p}}^{\alpha_{p} \beta_{p}} \Or_{\gamma_{p+1}}^{\alpha_{p+1}}, \\
\Obd_{(p)\alpha}^{\phantom{(p)}\gamma} &:=
\dfrac{1}{2^p}\delta_{\alpha_1 \beta_1 \ldots \alpha_{p} \beta_{p} \alpha_{p+1} \alpha}^{\gamma_1 \delta_1 \ldots . \gamma_{p} \delta_{p} \gamma_{p+1} \gamma} \Mr_{\gamma_1 \delta_1}^{\alpha_1 \beta_1} \ldots \Mr_{\gamma_{p} \delta_{p}}^{\alpha_{p} \beta_{p}} \Or_{\gamma_{p+1}}^{\alpha_{p+1}}, \\
\Ob_{(p)} &:=
\dfrac{1}{2^p}\delta_{\alpha_1 \beta_1 \ldots \alpha_{p} \beta_{p} \alpha_{p+1}}^{\gamma_1 \delta_1 \ldots . \gamma_{p} \delta_{p} \gamma_{p+1}} \Mr_{\gamma_1 \delta_1}^{\alpha_1 \beta_1} \ldots \Mr_{\gamma_{p} \delta_{p}}^{\alpha_{p} \beta_{p}} \Or_{\gamma_{p+1}}^{\alpha_{p+1}}, \\
\Or_{\alpha}^{\gamma} &=
-\Lc_u K_{\alpha}^{\gamma} - K_{\alpha \tau}K^{\tau \gamma} - \varepsilon \dfrac{\nabla_\alpha \nabla^\gamma N}{N}, \\
\bottomrule
\end{array}\]

All these tensors are defined on $\Mc$, so they are invariant under the action of $\gamma_\mu^\nu$ and vanish under the product with $u^\mu$. When they act on quantities defined on $\Mc$, ie. when the indices are latin, we shall indifferently write $\delta_i^j = \gamma_i^j = g_i^j$.

\subsection{Hamiltonian constraint}

Let us explicitly compute the projection of the hamiltonian constraint \eqref{eqcontrainte1fLL}. This term is quite easy to handle and uses the same algebraic operations as for the calculation of the momentum constraint \eqref{eqcontrainte2fLL} and the dynamical equations \eqref{eqdynfLL}.

We start from
\begin{align*}
\Rbd_{(p)}^{\mu\nu}u_\mu u_\nu
	= \Rbd_{(p)\mu}^{\phantom{(p)}\nu}u^\mu u_\nu
	= \dfrac{1}{2^p} \delta_{\alpha_1 \beta_1 \ldots \alpha_p \beta_p \mu}^{\gamma_1 \delta_1 \ldots . \gamma_p \delta_p \nu} &\Rr_{\gamma_1 \delta_1}^{\alpha_1 \beta_1} \Rr_{\gamma_2 \delta_2}^{\alpha_2 \beta_2} \ldots \Rr_{\gamma_p \delta_p}^{\alpha_p \beta_p} u^\mu u_\nu \\
	= \dfrac{1}{2^p} \delta_{\alpha_1 \beta_1 \ldots \alpha_p \beta_p \mu}^{\gamma_1 \delta_1 \ldots . \gamma_p \delta_p \nu}
	&\left(\Mr_{\gamma_1 \delta_1}^{\alpha_1 \beta_1} + \Nrb_{\gamma_1 \delta_1}^{\alpha_1 \beta_1} + \Orb_{\gamma_1 \delta_1}^{\alpha_1 \beta_1}\right) \\
	& \cdots \\
	&\left(\Mr_{\gamma_p \delta_p}^{\alpha_p \beta_p} + \Nrb_{\gamma_p \delta_p}^{\alpha_p \beta_p} + \Orb_{\gamma_p \delta_p}^{\alpha_p \beta_p}\right)u^\mu u_\nu.
\end{align*}
When we develop this product, we find terms containing arbitrary numbers of $\Mr_{\gamma \delta}^{\alpha \beta}$, $\Nr_{\gamma \delta}^{\alpha} u^{\beta}$, $\Nr^{\alpha \beta}_{\gamma} u_{\delta}$, and $\Or_{\gamma}^{\alpha} u^{\beta} u_{\delta}$. But two major facts cancel most of these terms:
\begin{itemize}
 \item when a term contains more than one $u^\alpha$ or $u_\alpha$, because of the antisymmetry of the determinant;
 \item when a term contains only contractions of $u_\alpha$ or $u^\gamma$ with $\Mr_{\gamma \delta}^{\alpha \beta}$, $\Nr_{\gamma \delta}^{\alpha}$, $O_\gamma^\alpha$, which are by definition orthogonal to $u_\alpha$ and $u^\gamma$.
\end{itemize}
At last, we are left with the only term in which $u^\mu$ hits $u_\nu$, that is to say
\begin{align*}
\Rbd_{(p)}^{\mu\nu}u_\mu u_\nu
	= \dfrac{1}{2^p} \delta_{\alpha_1 \beta_1 \ldots \alpha_p \beta_p \mu}^{\gamma_1 \delta_1 \ldots \gamma_p \delta_p \nu} \Mr_{\gamma_1 \delta_1}^{\alpha_1 \beta_1} \Mr_{\gamma_2 \delta_2}^{\alpha_2 \beta_2} \ldots \Mr_{\gamma_p \delta_p}^{\alpha_p \beta_p} u^\mu u_\nu.
\end{align*}
While we expand along the last column, all the terms vanish except
\begin{align}
\Rbd_{(p)}^{\mu\nu}u_\mu u_\nu
	&= \dfrac{1}{2^p} \delta_{\alpha_1 \beta_1 \ldots \alpha_p \beta_p}^{\gamma_1 \delta_1 \ldots \gamma_p \delta_p} \delta_\mu^\nu \Mr_{\gamma_1 \delta_1}^{\alpha_1 \beta_1} \Mr_{\gamma_2 \delta_2}^{\alpha_2 \beta_2} \ldots \Mr_{\gamma_p \delta_p}^{\alpha_p \beta_p} u^\mu u_\nu \nonumber \\
	&= \varepsilon \left[\dfrac{1}{2^p} \delta_{\alpha_1 \beta_1 \ldots \alpha_p \beta_p}^{\gamma_1 \delta_1 \ldots \gamma_p \delta_p} \Mr_{\gamma_1 \delta_1}^{\alpha_1 \beta_1} \Mr_{\gamma_2 \delta_2}^{\alpha_2 \beta_2} \ldots \Mr_{\gamma_p \delta_p}^{\alpha_p \beta_p}\right] \nonumber \\
	&= \varepsilon \Mb_{(p)}. \label{Rbduu}
\end{align}

The second projection to compute is, according to the definition \eqref{defP},
\begin{align*}
P_{(p)}^{\mu\alpha\beta\nu}\phi_{p|\alpha\beta}u_\mu u_\nu
  &= \dfrac{1}{2}\Rbdd_{(p-1)\sigma\tau}^{\phantom{(p-1)}\mu\alpha} g^{\sigma \beta} g^{\tau \nu}\phi_{p|\alpha\beta}u_\mu u_\nu \\
  &= \dfrac{1}{2^p}\delta_{\alpha_1 \beta_1 \ldots \alpha_{p-1} \beta_{p-1} \sigma \tau}^{\gamma_1 \delta_1 .\ldots \gamma_{p-1} \delta_{p-1} \mu \alpha} \Rr_{\gamma_1 \delta_1}^{\alpha_1 \beta_1} \ldots \Rr_{\gamma_{p-1} \delta_{p-1}}^{\alpha_{p-1} \beta_{p-1}}
  g^{\sigma \beta} g^{\tau \nu}\phi_{p|\alpha\beta}u_\mu u_\nu \\
  &= -\dfrac{1}{2^p}\delta_{\alpha_1 \beta_1 \ldots \alpha_{p-1} \beta_{p-1} \beta \nu}^{\gamma_1 \delta_1 .\ldots \gamma_{p-1} \delta_{p-1} \alpha \mu} \Rr_{\gamma_1 \delta_1}^{\alpha_1 \beta_1} \ldots \Rr_{\gamma_{p-1} \delta_{p-1}}^{\alpha_{p-1} \beta_{p-1}}
  \phi_{p|\alpha}^{\phantom{p}|\beta} u_\mu u^\nu.
\end{align*}
As previously, the presence of $u_\mu$ and $u^\nu$ cancels all the terms containing at least one among $\Nr_{\gamma \delta}^{\alpha} u^{\beta}$, $\Nr^{\alpha \beta}_{\gamma} u_{\delta}$, or $\Or_{\gamma}^{\alpha} u^{\beta} u_{\delta}$. Hence,
\begin{align}
P_{(p)}^{\mu\alpha\beta\nu}\phi_{p|\alpha\beta}u_\mu u_\nu
  &= -\dfrac{1}{2^p}\delta_{\alpha_1 \beta_1 \ldots \alpha_{p-1} \beta_{p-1} \beta \nu}^{\gamma_1 \delta_1 .\ldots \gamma_{p-1} \delta_{p-1} \alpha \mu} \Mr_{\gamma_1 \delta_1}^{\alpha_1 \beta_1} \ldots \Mr_{\gamma_{p-1} \delta_{p-1}}^{\alpha_{p-1} \beta_{p-1}}
  \phi_{p|\alpha}^{\phantom{p}|\beta} u_\mu u^\nu.
\end{align}
Thereafter we expand along the last row:
\begin{align}
P_{(p)}^{\mu\alpha\beta\nu}\phi_{p|\alpha\beta}u_\mu u_\nu
  &= -\dfrac{1}{2^p}\bigg[\delta_{\alpha_1 \beta_1 \ldots \alpha_{p-1} \beta_{p-1} \beta}^{\gamma_1 \delta_1 .\ldots \gamma_{p-1} \delta_{p-1} \alpha} \delta_\nu^\mu
  \Mr_{\gamma_1 \delta_1}^{\alpha_1 \beta_1} \ldots \Mr_{\gamma_{p-1} \delta_{p-1}}^{\alpha_{p-1} \beta_{p-1}}
  u_\mu u^\nu \nonumber \\
  &\hspace{15mm} -\delta_{\alpha_1 \beta_1 \ldots \alpha_{p-1} \beta_{p-1} \beta}^{\gamma_1 \delta_1 .\ldots \gamma_{p-1} \delta_{p-1} \mu} \delta_\nu^\alpha
  \Mr_{\gamma_1 \delta_1}^{\alpha_1 \beta_1} \ldots \Mr_{\gamma_{p-1} \delta_{p-1}}^{\alpha_{p-1} \beta_{p-1}} u_\mu u^\nu
  \bigg] \phi_{p|\alpha}^{\phantom{p}|\beta} \nonumber \\
  &= -\dfrac{1}{2^p} \bigg[ \varepsilon \delta_{\alpha_1 \beta_1 \ldots \alpha_{p-1} \beta_{p-1} \beta}^{\gamma_1 \delta_1 .\ldots \gamma_{p-1} \delta_{p-1} \alpha}
  \Mr_{\gamma_1 \delta_1}^{\alpha_1 \beta_1} \ldots \Mr_{\gamma_{p-1} \delta_{p-1}}^{\alpha_{p-1} \beta_{p-1}} \nonumber \\
  &\hspace{15mm} -\delta_{\alpha_1 \beta_1 \ldots \alpha_{p-1} \beta_{p-1}}^{\gamma_1 \delta_1 .\ldots \gamma_{p-1} \delta_{p-1}} \delta_\beta^\mu
  \Mr_{\gamma_1 \delta_1}^{\alpha_1 \beta_1} \ldots \Mr_{\gamma_{p-1} \delta_{p-1}}^{\alpha_{p-1} \beta_{p-1}} u_\mu u^\alpha \bigg] \phi_{p|\alpha}^{\phantom{p}|\beta} \nonumber \\
  &= -\dfrac{1}{2^p}\bigg[
  \delta_{\alpha_1 \beta_1 \ldots \alpha_{p-1} \beta_{p-1}}^{\gamma_1 \delta_1 .\ldots \gamma_{p-1} \delta_{p-1}}
  \Mr_{\gamma_1 \delta_1}^{\alpha_1 \beta_1} \ldots \Mr_{\gamma_{p-1} \delta_{p-1}}^{\alpha_{p-1} \beta_{p-1}}
  \left(\varepsilon\delta_\beta^\alpha - u_\beta u^\alpha\right) \nonumber \\
  & \hspace{15mm} - \varepsilon 2(p-1) \delta_{\alpha_1 \beta_1 \ldots \alpha_{p-1} \beta}^{\gamma_1 \delta_1 .\ldots \gamma_{p-1} \delta_{p-1}} \delta_{\beta_{p-1}}^\alpha
  \Mr_{\gamma_1 \delta_1}^{\alpha_1 \beta_1} \ldots \Mr_{\gamma_{p-1} \delta_{p-1}}^{\alpha_{p-1} \beta_{p-1}} \bigg] \phi_{p|\alpha}^{\phantom{p}|\beta} \nonumber \\
  &= \bigg[-\dfrac{\varepsilon}{2} \Mb_{(p-1)} \gamma_\beta^\alpha \nonumber \\
  & \hspace{15mm} + \dfrac{\varepsilon(p-1)}{2}\left(\dfrac{1}{2^{p-2}}
  \delta_{\alpha_1 \beta_1 \ldots \alpha_{p-2} \beta_{p-2} k \beta}^{\gamma_1 \delta_1 .\ldots \gamma_{p-2} \delta_{p-2} i j}
\Mr_{\gamma_1 \delta_1}^{\alpha_1 \beta_1} \ldots \Mr_{\gamma_{p-2} \delta_{p-2}}^{\alpha_{p-2} \beta_{p-2}} \right)
  \Mr_{i j}^{k \alpha} \bigg] \phi_{p|\alpha}^{\phantom{p}|\beta} \nonumber \\
  &= \bigg[-\dfrac{\varepsilon}{2} \Mb_{(p-1)} \gamma_\beta^\alpha
  + \dfrac{\varepsilon(p-1)}{2} \Mbdd_{(p-2)k\beta}^{\phantom{(p-2)}ij} \Mr_{i j}^{k \alpha} \bigg] \phi_{p|\alpha}^{\phantom{p}|\beta} \nonumber \\
  &= \bigg[-\dfrac{\varepsilon}{2} \Mb_{(p-1)} \gamma_b^a
  + \dfrac{\varepsilon(p-1)}{2} \Mbdd_{(p-2)k b}^{\phantom{(p-2)}i j} \Mr_{i j}^{k a} \bigg] \left(\gamma_a^\alpha \gamma_\beta^b \phi_{p|\alpha}^{\phantom{p}|\beta}\right). \label{Puu}
\end{align}

So,
\begin{align}
\Ac_{(p)}^{\mu\nu}u_\mu u_\nu
  &= \left[\phi_p\Rbd_{(p)}^{\mu\nu} +2p\left(P_{(p)}^{\mu\rho\sigma\nu} + P_{(p)}^{\mu\sigma\rho\nu}\right)\phi_{p|\rho\sigma}\right] u_\mu u_\nu \nonumber \\
	&= \varepsilon \left[\phi_p \Mb_{(p)} - 2p \left(\Mb_{(p-1)} \gamma_b^a - (p-1) \Mbdd_{(p-2)k b}^{\phantom{(p-2)}i j} \Mr_{i j}^{k m}\right) \left(\gamma_a^\alpha \gamma_\beta^b \phi_{p|\alpha}^{\phantom{p}|\beta}\right)\right] \nonumber \\
	&= \varepsilon \left[\phi_p \Mb_{(p)} - 2p \left(\Mb_{(p-1)} \gamma_b^a - (p-1) \Mbdd_{(p-2)k b}^{\phantom{(p-2)}i j} \Mr_{i j}^{k a}\right) \left(\nabla_a \nabla^b \phi_p + \varepsilon \Pi_p K_a^b \right)\right], \label{Auu}
\end{align}
according to \eqref{ggDDphi}.

The computations of the momentum constraint and the dynamical equations, more involved but not more difficult, use the same methods. They are left in \nameref{appendix}.

\section{\texorpdfstring{$f(\text{Lovelock})$}{f(Lovelock)} decomposition: results and applications}

Let us gather here the results from section \ref{sectionoutline} and \nameref{appendix}.

\subsection{Results}
We have just done the hamiltonian decomposition of $f(\text{Lovelock})$ gravity.

The hamiltonian constraint \eqref{eqcontrainte1fLL} writes
\begin{align}
\varepsilon V - E
	&= \sum_{p=0}^{p_n} \Ac_{(p)}^{\mu\nu}u_\mu u_\nu \nonumber \\
	&= \sum_{p=0}^{p_n} \varepsilon \left[\phi_p \Mb_{(p)} - 2p \left(\Mb_{(p-1)} \gamma_b^a - (p-1) \Mbdd_{(p-2)k b}^{\phantom{(p-2)}i j} \Mr_{i j}^{k a}\right) \left(\nabla_a \nabla^b \phi_p + \varepsilon \Pi_p K_a^b \right)\right]. \label{eqcontrainte1fLLcomplet}
\end{align}

The momentum constraint \eqref{eqcontrainte2fLL} is
\begin{align}
-\varepsilon J^i
	=~& \sum_{p=0}^{p_n} \Ac_{(p)}^{\mu\nu}\gamma_{\mu}^i u_\nu \nonumber \\
	=~& \sum_{p=0}^{p_n} \phi_p\left(-\varepsilon 2p \Nb_{(p)}^{\phantom{(p)}i}\right) \nonumber \\
	& + 2\varepsilon p^2 \bigg[ 2 \Nb_{(p-1)}^{\phantom{(p-1)}i} \gamma_b^a - 2 \Nb_{(p-1)}^{\phantom{(p-1)}a} \gamma_b^i \nonumber \\
	& \hspace{15mm} + \Mbdd_{(p-2)k l}^{\phantom{(p-2)} i a} \Nr_b^{k l} - 2(p-2) \Nbdd_{(p-2)\phantom{i} k l}^{\phantom{(p-2)}i j a} \Mr_{j b}^{k l} \bigg] \left(\nabla_a \nabla^b \phi_p + \varepsilon \Pi_p K_a^b \right) \nonumber \\
	& + 2p \Mbd_{(p-1) b}^{\phantom{(p-1)} i} \left(\nabla^b \Pi_p - K^{ab}\nabla_a \phi_p \right). \label{eqcontrainte2fLLcomplet}
\end{align}

And the dynamical equations \eqref{eqdynfLL} are
\begin{align}
	&\hspace{20mm}
	V \gamma_j^i - S_j^i = \sum_{p=0}^{p_n} \Ac_{(p)}^{\mu\nu}\gamma_{\mu}^i \gamma_{j\nu} \nonumber \\
	&= \sum_{p=0}^{p_n} \left[\Mbd_{(p)j}^{\phantom{(p)}i} + \varepsilon 2p(p-1)\NNbd_{(p-1)j}^{\phantom{(p-1)}i} + \varepsilon 2p\Obd_{(p-1)j}^{\phantom{(p-1)}i}\right]\phi_p \nonumber \\
	& -p \bigg[\Mbdd_{(p-1)b j}^{\phantom{(p-1)} a i} + 2\varepsilon (p-1)(p-2) \NNbdd_{(p-2) b j}^{\phantom{(p-2)} a i} + 2\varepsilon (p-1)\Obdd_{(p-2) b j}^{\phantom{(p-2)} a i} \bigg]
	\left(\nabla_a \nabla^b \phi_p + \varepsilon \Pi_p K_a^b \right) \nonumber \\
	&+ 2p(p-1) \left[
	- \Nb_{(p-1)}^{\phantom{(p-1)}i} \delta_j^a
	+ \Mbdd_{(p-2)l j}^{\phantom{(p-2)} k i} \Nr_{k}^{l \, a}
	+ (p-2) \Nbdd_{(p-2) \phantom{i} k j}^{\phantom{(p-2)} i c d} \Mr_{c d}^{k a} \right] \left(\nabla_a \Pi_p - K_a^b \nabla_b \phi_p\right) \nonumber \\
	&+ 2p(p-1) \left[
	- \Nb_{(p-1)j} \delta_b^i
	+ \Mbdd_{(p-2)l j}^{\phantom{(p-2)} k i} \Nr_{k b}^{l}
	+ (p-2) \Nbdd_{(p-2) j c d}^{\phantom{(p-2) j} k i} \Mr_{k b}^{c d} \right] \left(\nabla^b \Pi_p - K_a^b \nabla^a \phi_p\right) \nonumber \\
	&-\varepsilon p \Mbd_{(p-1)j}^{\phantom{(p-1)}i} \big(\varepsilon\left(\square \phi_p - \Delta \phi_p\right) - \Pi_p K \big). \label{eqdynfLLcomplet}
\end{align}


\subsection{Applications}
Let us look at small values of $p$, in order to recover some known cases.
\[\begin{array}{ll|ll}
\toprule
\Mb_{(-1)} &= 0 &
\Nb_{(-1)i}&= 0 \\
\Mb_{(0)}  &= 1 &
\Nb_{(0)i} &= 0 \\
\Mb_{(1)}  &= \Mr &
\Nb_{(1)i} &= \Nr_i \\
\Mb_{(2)}  &= \Mr^2 - 4 \Mr_{ij}\Mr^{ij} + \Mr_{ij}^{kl}\Mr_{kl}^{ij} &
\Nb_{(2)i} &= \Mr \Nr_i - 2\Mr_i^j \Nr_j - 2\Mr_j^k \Nr_{ki}^j + \Mr_{li}^{jk}\Nr_{jk}^l \\
\midrule
\Mbd_{(-1)j}^{\phantom{(-1)}i} &= 0 &
\Mbd_{(1)j}^{\phantom{(1)}i} &= \Mr \delta_j^i - 2\Mr_j^i \\
\Mbd_{(0)j}^{\phantom{(0)}i} &= \delta_j^i &
\Mbd_{(2)j}^{\phantom{(2)}i} &= \Mb_{(2)} \delta_j^i - 8\left(\Mr \Mr_j^i -2\Mr_k^l \Mr_{l j}^{k i}\right) \\
\midrule
\Mbdd_{(-2)ab}^{\phantom{(-2)}cd} &= 0 &
\Nbdd_{(-2)iab}^{\phantom{(-2)i}cd} &= 0 \\
\Mbdd_{(-1)ab}^{\phantom{(-1)}cd} &= 0 &
\Nbdd_{(-1)iab}^{\phantom{(-1)i}cd} &= 0 \\
\Mbdd_{(0)ab}^{\phantom{(0)}cd} &= \delta_a^c \delta_b^d - \delta_a^d \delta_b^c &
\Nbdd_{(0)iab}^{\phantom{(0)i}cd} &= 0 \\
\Mbdd_{(1)ab}^{\phantom{(1)}cd} &= \Mr \left(\delta_a^c \delta_b^d - \delta_a^d \delta_b^c\right)  + 2\Mr_{a b}^{c d} &
\Nbdd_{(1)iab}^{\phantom{(1)i}cd} &= \Nr_i \left(\delta_a^c \delta_b^d - \delta_a^d \delta_b^c\right) + \Nr_{ab}^c \delta_i^d - \Nr_{ab}^d \delta_i^c \\
	& - 2 \left(\Mr_a^c \delta_b^d + \Mr_b^d \delta_a^c - \Mr_a^d \delta_b^c - \Mr_b^c \delta_a^d\right)
	& & -\Nr_a \left(\delta_i^c \delta_b^d - \delta_i^d \delta_b^c\right)
		+\Nr_b \left(\delta_i^c \delta_a^d - \delta_i^d \delta_a^c\right) \\
	& & & -\left(\Nr_{ai}^c \delta_b^d + \Nr_{bi}^d \delta_a^c - \Nr_{ai}^d \delta_b^c - \Nr_{bi}^c \delta_a^d \right) \\
\midrule
\Obd_{(-1)j}^{\phantom{(-1)}i} &= 0 &
\NNbd_{(-1)j}^{\phantom{(-1)}i} &= 0 \\
\Obd_{(0)j}^{\phantom{(0)}i} &= \Or \delta_j^i - \Or_j^i &
\NNbd_{(0)j}^{\phantom{(0)}i} &= 0 \\
\Obd_{(1)j}^{\phantom{(1)}i} &= \left(\Mr \Or - 2\Mr_k^l \Or_l^k\right)\delta_j^i &
\NNbd_{(1)j}^{\phantom{(1)}i} &= -\left(\Nr_{a b}^c \Nr_c^{a b} + 2 \Nr_a \Nr^a\right) \delta_j^i \\
	& - \left(\Mr \Or_j^i - 2 \Mr_j^i \Or\right) &
	& + \Nr_a \Nr_j^{i a} + 2\Nr_{a b}^i \Nr_j^{a b} \\
	& + 2\left(\Mr_k^i \Or_j^k + \Mr_j^k \Or_k^i + \Mr_{k j}^{l i} \Or_l^k\right) &
	& + 2\left(\Nr_j \Nr^i - \Nr_{a j}^c \Nr_c^{a i} + \Nr_{j a}^i \Nr^a\right) \\
\midrule
\Obdd_{(-2)ab}^{\phantom{(-1)}cd} &= 0 &
\NNbdd_{(-2)ab}^{\phantom{(-1)}cd} &= 0 \\
\Obdd_{(-1)ab}^{\phantom{(-1)}cd} &= 0 &
\NNbdd_{(-1)ab}^{\phantom{(-1)}cd} &= 0 \\
\Obdd_{(0)ab}^{\phantom{(0)}cd} &= \Or \left(\delta_a^c \delta_b^d - \delta_a^d \delta_b^c\right) &
\NNbdd_{(0)ab}^{\phantom{(0)}cd} &= 0 \\
	& - \left(\Or_a^c \delta_b^d + \Or_b^d \delta_a^c - \Or_b^c \delta_a^d - \Or_a^d \delta_b^c\right) \\
\bottomrule
\end{array}\]
To get these expressions, we can use formulas like
\begin{equation}
\Mbdd_{(p)ab}^{\phantom{(p)}cd} = \left(\Mbd_{(p)a}^{\phantom{(p)} c} \delta_b^d - \Mbd_{(p)a}^{\phantom{(p)} d} \delta_b^c \right) + p\left(\Mbdd_{(p-1)i j}^{\phantom{(p-1)}k c} \Mr_{k b}^{i j} \delta_a^d - 2p\Mbdd_{(p-1)i a}^{\phantom{(p-1)}k c} \Mr_{k b}^{i d}\right),
\end{equation}
which follow from successive developments along rows.

Then for $p=0$ we get
\begin{align}
\Ac_{(0)}^{\mu\nu} u_\mu u_\nu
	&= \varepsilon \phi_0, \\
\Ac_{(0)}^{\mu\nu} \gamma_{\mu}^i u_\nu
	&= 0, \\
\Ac_{(0)}^{\mu\nu} \gamma_{\mu}^i \gamma_{j\nu}
	&= \phi_0 \delta_j^i.
\end{align}

For $p=1$, we obtain
\begin{align}
\Ac_{(1)}^{\mu\nu} u_\mu u_\nu
	&= \varepsilon \left[\phi_1 \Mr - 2 \left(\Delta \phi_1 + \varepsilon \Pi_1 K\right)\right], \\
\Ac_{(1)}^{\mu\nu} \gamma_{\mu}^i u_\nu
	&= -2\varepsilon \phi_1 \Nr^i, \\
\Ac_{(1)}^{\mu\nu} \gamma_{\mu}^i \gamma_{j\nu}
	&= \phi_1 \left[\left(\Mr \delta_j^i - 2\Mr_j^i\right) + 2\varepsilon \left(\Or \delta_j^i - \Or_j^i\right)\right] \nonumber \\
	& - \delta_j^i \square \phi_1 + \nabla_j \nabla^i \phi_1 + \varepsilon \Pi_1 K_j^i.
\end{align}
This corresponds to $f(\Rr)$ theories (see for instance \cite{Sot10}). If $\phi_1 \equiv 1$, it is nothing but GR.

For $p=2$, we get
\begin{align}
\Ac_{(2)}^{\mu\nu} u_\mu u_\nu
	&= \varepsilon \phi_2 \left(\Mr^2 - 4 \Mr_{ij}\Mr^{ij} + \Mr_{ij}^{kl}\Mr_{kl}^{ij}\right) \nonumber \\
	&- 4 \varepsilon\left(\Mr \delta_b^a - 2 \Mr_b^a\right) \left(\nabla_a \nabla^b \phi_2 + \varepsilon \Pi_2 K_a^b \right), \label{fRGB1} \\
\Ac_{(2)}^{\mu\nu} \gamma_{\mu}^i u_\nu
	&= -4\varepsilon \phi_2 \left(\Mr \Nr^i - 2\Mr_j^i \Nr^j - 2\Mr_j^k \Nr_{k}^{ji} + \Mr_{jk}^{li} \Nr_l^{jk}\right) \nonumber \\
	& + 16\varepsilon \left(\Nr^i \delta_b^a - \Nr^a \delta_b^i - \Nr_b^{a i}\right) \left(\nabla_a \nabla^b \phi_2 + \varepsilon \Pi_2 K_a^b \right) \nonumber \\
	& + 4 \left(\Mr \delta_b^i - 2 \Mr_b^i\right) \left(\nabla^b \Pi_2 - K^{ab}\nabla_a \phi_2 \right),
	\label{fRGB2} \\
\Ac_{(2)}^{\mu\nu} \gamma_{\mu}^i \gamma_{j\nu}
	&= \phi_2 \bigg[
	\left(\left(\Mr^2 - 4 \Mr_{ij}\Mr^{ij} + \Mr_{ij}^{kl}\Mr_{kl}^{ij}\right) \delta_j^i - 8\left(\Mr \Mr_j^i -2\Mr_k^l \Mr_{l j}^{k i}\right)\right) \nonumber \\
	&+ 4 \varepsilon \left( -\left(\Nr_{a b}^c \Nr_c^{a b} + 2 \Nr_a \Nr^a\right) \delta_j^i
	+ \Nr_a \Nr_j^{i a} + 2\Nr_{a b}^i \Nr_j^{a b}
	+ 2\left(\Nr_j \Nr^i - \Nr_{a j}^c \Nr_c^{a i} + \Nr_{j a}^i \Nr^a\right)\right) \nonumber \\
	&+ 4 \varepsilon \left(\left(\Mr \Or - 2\Mr_k^l \Or_l^k\right)\delta_j^i
	- \left(\Mr \Or_j^i - 2 \Mr_j^i \Or\right)
	+ 2\left(\Mr_k^i \Or_j^k + \Mr_j^k \Or_k^i + \Mr_{k j}^{l i} \Or_l^k\right)\right) \bigg] \nonumber \\
	&- 2\left[\Mr \delta_j^i - 2\Mr_j^i\right] \square \phi_2 \nonumber \\
	&+ 2\bigg[\Mr \delta_j^a \delta_b^i - 2\Mr_j^a \delta_b^i - 2 \Mr_b^i \delta_j^a + 2 \Mr_b^a \delta_j^i - 2 \Mr_{j b}^{i a} \nonumber \\
	&- 2\varepsilon \left(\Or \left(\delta_b^a \delta_j^i - \delta_b^i \delta_j^a\right) - \left(\Or_b^a \delta_j^i + \Or_j^i \delta_b^a - \Or_b^i \delta_j^a - \Or_j^a \delta_b^i\right)\right)\bigg]
	\left(\nabla_a \nabla^b \phi_2 + \varepsilon \Pi_2 K_a^b \right) \nonumber \\
	&+ 4 \left[- \Nr^i \delta_j^a + \Nr^a \delta_j^i - \Nr_j^{i a}
	\right]\left(\nabla_a \Pi_2 - K_a^b \nabla_b \phi_2 \right) \nonumber \\
	&+ 4 \left[- \Nr_j \delta_b^i + \Nr_b \delta_j^i - \Nr_{j b}^i
	\right]\left(\nabla^b \Pi_2 - K_a^b \nabla^a \phi_2 \right). \label{fRGB3}
\end{align}
When $\phi_2$ is constant, we recover the Gauss-Bonnet gravity (see for instance \cite{Tor08}), $\Rb_2$ being the Gauss-Bonnet term for $n+1=4$. The equations for a non-constant $\phi_2$ did not seem to be known previously.

More generally, although the cases $p \leq 1$ and $\forall \, p, \, \phi_p = \text{cst.}$ were already known (see \cite{Tei87} and \cite{Cho88}), it seems to us that the general equations $(\ref{eqcontrainte1fLLcomplet},\ref{eqcontrainte2fLLcomplet},\ref{eqdynfLLcomplet})$ had not been made explicit up to now.

\section{Conclusion}
In this paper, we have derived the constraint and dynamical equations of $f(\text{Lovelock})$ theories. This family of modified gravity theories appears naturally as the common generalisation of $f(\Rr)$ and Lovelock theories, and could bring interesting results in cosmology or string/M-theories. Only the particular case of $f(\Rr$, Gauss-Bonnet$)$ was studied until the paper of Bueno et al. \cite{Bueno16}, which considers the general $f($Lovelock$)$ case.

The field equations of the $f(\text{Lovelock})$ theories were already presented in \cite{Bueno16}, but here we explain how they are obtained. We write the action in a scalar-tensor shape thanks to a Legendre transform, when the hessian of $f$ is invertible, and then we use a theorem of H. Rund that gives the expression of the total derivative of a lagrangian density with respect to the metric.

Once we have the field equations, we project them on a space-like hypersurface and on its normal unit vector. This gives three sets of equations: the hamiltonian constraint, the momentum constraint and the dynamical equations. The field equations contain highly non-linear terms; but we invoke the properties of the determinant and expand several times those terms. We can express the non-linear terms from the orthogonal decomposition of the Riemann tensor, and most of them cancel for orthogonality or antisymmetry reasons. This enables us to give the $n+1$ decomposition of $f(\text{Lovelock})$ field equations solely in terms of data on the hypersurface and their derivatives. This is the system $(\ref{eqcontrainte1fLLcomplet}, \ref{eqcontrainte2fLLcomplet}, \ref{eqdynfLLcomplet})$. It involves $(\Mb, \Mbd, \Mbdd, \Nb, \Nbdd, \NNbd, \NNbdd, \Obd, \Obdd)$ which are only products of the projections of the Riemann tensor $(\Mr_{ij}^{kl}, \Nr_{ij}^{kl}, \Or_{ij}^{kl})$, which in turn are expressed from $(\gamma_{ij}, K_{ij})$; and from $(\phi_p, \Pi_p)$. Thus we obtain the same hamiltonian structure as for GR.

We test our formulas for $p \leq 1$ and recover the known $n+1$ decomposition of $f(\Rr)$ gravity. For $p=2$ and $\phi_2$ constant, we recover the explicit results of \cite{Tor08}. For $p$ arbitrary and $\phi_p$ constant, we recover the results of \cite{Tei87} and \cite{Cho88}.

All the other cases seemed to be unknown before, hence we produce original expressions for $f(\text{Lovelock})$ theories: the constraint $(\ref{eqcontrainte1fLLcomplet}, \ref{eqcontrainte2fLLcomplet})$ and the dynamical $(\ref{eqdynfLLcomplet})$ equations. The restriction to $p=2$ gives the explicit decomposition for $f(\Rr, \text{Gauss-Bonnet})$ gravity: $(\ref{fRGB1}, \ref{fRGB2}, \ref{fRGB3})$.

\vspace{\baselineskip}

The well-posedness of the $f(\text{Lovelock})$ gravity is still an unexplored field of research.

If we put aside the evolution problem, ie. the dynamical equations \eqref{eqdynfLLcomplet} of $f(\text{Lovelock})$ theories and their well-posedness, we are left with the constraint equations $(\ref{eqcontrainte1fLLcomplet},\ref{eqcontrainte2fLLcomplet})$. The GR constraint equations are an interesting and fertile source of mathematical research, and it is worth it to ask whether the results applying to GR constraint equations are still valid for $f(\text{Lovelock})$ constraint equations $(\ref{eqcontrainte1fLLcomplet},\ref{eqcontrainte2fLLcomplet})$.

The GR constraint equations are underdetermined. The usual way to solve them is the conformal method: to search a solution metric in a given conformal class. It is thus natural to seek to adapt this method to $(\ref{eqcontrainte1fLLcomplet}, \ref{eqcontrainte2fLLcomplet})$. Let us look at the most simple case: the Lovelock theories, with a time-symmetric ADM-decomposition.
\[\forall \, p,\ \phi_p = \text{cst.} \qquad \text{and} \qquad K_{ij} = 0.\]
Then the momentum constraint reduces to $J^i = 0$, which is a condition on the stress-energy tensor so that a time-symmetric decomposition be possible. The hamiltonian constraint becomes
\begin{equation}
\sum_{p=0}^{p_n} \phi_p \Rb_{p}^\Mc = V - \varepsilon E. \label{eqcontrainte1LLtsym}
\end{equation}
It is a new geometrical equation. When all but one of the $\phi_p$'s are taken to be $0$, $(\ref{eqcontrainte1LLtsym})$ is only a curvature prescription equation. Hence, searching a solution of this equation in a given conformal class is a generalised Yamabe problem. In the conformally flat case, it can be shown that $\Rb_{p}^\Mc$ is nothing but the $\sigma_k$-curvature of $\Mc$ (see for example \cite{Ge14}). So $(\ref{eqcontrainte1LLtsym})$ is a problem of $\sigma_k$-curvature prescription: the Lovelock products fulfil their promise to raise geometrical interests. This connection between Lovelock theories and the $\sigma_k$-curvatures was first done in \cite{Lab08}.

The $\sigma_k$-Yamabe problem, ie. the search for a conformal factor for which a $\sigma_k$-curvature is constant, had been solved in the $2000$'s (see \cite{Lab08} for references). We handle for the first time in \cite{Lac17-2} the Lovelock case, where more than one $\phi_p$ are not zero, ie. the prescription problem for an arbitrary linear combination of $\sigma_k$-curvatures.

The mathematical resolution of the general $f(\text{Lovelock})$ constraint equations, without any assumptions, is entirely open.

\section*{Acknowledgements}
I would like to deeply thank Dr. Loïc Villain, who introduced me to this area of research.

\section*{Appendix} \label{appendix}

\subsection*{Momentum constraint}
\begin{align*}
\Rbd_{(p)}^{\mu\nu}\gamma_{i\mu} u_\nu
	= \Rbd_{(p)\mu}^{\phantom{(p)}\nu}\gamma_i^\mu u_\nu \\
	= \dfrac{1}{2^p} \delta_{\alpha_1 \beta_1 \ldots \alpha_p \beta_p \mu}^{\gamma_1 \delta_1 \ldots . \gamma_p \delta_p \nu}
	&\Rr_{\gamma_1 \delta_1}^{\alpha_1 \beta_1} \Rr_{\gamma_2 \delta_2}^{\alpha_2 \beta_2} \ldots \Rr_{\gamma_p \delta_p}^{\alpha_p \beta_p}\gamma_i^\mu u_\nu \\
	= \dfrac{1}{2^p} \delta_{\alpha_1 \beta_1 \ldots \alpha_p \beta_p \mu}^{\gamma_1 \delta_1 \ldots . \gamma_p \delta_p \nu}
	&\left(\Mr_{\gamma_1 \delta_1}^{\alpha_1 \beta_1} + \Nrb_{\gamma_1 \delta_1}^{\alpha_1 \beta_1} + \Orb_{\gamma_1 \delta_1}^{\alpha_1 \beta_1}\right) \\
	& \cdots \\
	&\left(\Mr_{\gamma_p \delta_p}^{\alpha_p \beta_p} + \Nrb_{\gamma_p \delta_p}^{\alpha_p \beta_p} + \Orb_{\gamma_p \delta_p}^{\alpha_p \beta_p}\right)\gamma_i^\mu u_\nu.
\end{align*}
In the developed product, the terms containing at least one $\Or_{\gamma}^{\alpha} u_\delta u^\beta$ or one $\Nr_{\gamma}^{\alpha\beta} u_\delta$ vanish, because of the presence of $u_\nu$. The same antisymmetry cancels as well the terms with more than one $\Nr_{\gamma\delta}^{\alpha} u^\beta$, and the orthogonality of $u_\nu$ with $\gamma_i^\mu$ and $\Mr_{\gamma\delta}^{\alpha\beta}$ cancels the term containing only $\Mr_{\gamma\delta}^{\alpha\beta}$. So we only keep
\begin{align}
\Rbd_{(p)}^{\mu\nu}\gamma_{i\mu} u_\nu
	&= \dfrac{2p}{2^p} \delta_{\alpha_1 \beta_1 \ldots \alpha_p \beta_p \mu}^{\gamma_1 \delta_1 \ldots . \gamma_p \delta_p \nu}
	\Mr_{\gamma_1 \delta_1}^{\alpha_1 \beta_1} \Mr_{\gamma_2 \delta_2}^{\alpha_2 \beta_2} \ldots \Nr_{\gamma_p \delta_p}^{\alpha_p} u^{\beta_p} \gamma_i^\mu u_\nu \nonumber \\
	&= -\dfrac{2p}{2^p} \delta_{\alpha_1 \beta_1 \ldots \alpha_p \mu}^{\gamma_1 \delta_1 \ldots . \gamma_p \delta_p}\delta_{\beta_p}^\nu
	\Mr_{\gamma_1 \delta_1}^{\alpha_1 \beta_1} \Mr_{\gamma_2 \delta_2}^{\alpha_2 \beta_2} \ldots \Nr_{\gamma_p \delta_p}^{\alpha_p} u^{\beta_p} \gamma_i^\mu u_\nu \nonumber \\
	&= -\varepsilon\dfrac{2p}{2^p} \delta_{\alpha_1 \beta_1 \ldots \alpha_p \mu}^{\gamma_1 \delta_1 \ldots \gamma_p \delta_p}
	\Mr_{\gamma_1 \delta_1}^{\alpha_1 \beta_1} \Mr_{\gamma_2 \delta_2}^{\alpha_2 \beta_2} \ldots \Nr_{\gamma_p \delta_p}^{\alpha_p} \gamma_i^\mu \nonumber \\
	&= -\varepsilon 2p \Nb_{(p)i}. \label{Rbdgu}
\end{align}

Now the second projection:
\begin{align}
P_{(p)}^{\mu\alpha\beta\nu}\phi_{p|\alpha\beta}\gamma_{i\mu} u_\nu
  &= \dfrac{1}{2}\Rbdd_{(p-1)\sigma\tau}^{\phantom{(p-1)}\mu\alpha} g^{\sigma \beta} g^{\tau \nu}\phi_{p|\alpha\beta}\gamma_{i\mu} u_\nu \nonumber \\
  &= \dfrac{1}{2^p}\delta_{\alpha_1 \beta_1 \ldots \alpha_{p-1} \beta_{p-1} \sigma \tau}^{\gamma_1 \delta_1 .\ldots \gamma_{p-1} \delta_{p-1} \mu \alpha} \Rr_{\gamma_1 \delta_1}^{\alpha_1 \beta_1} \ldots \Rr_{\gamma_{p-1} \delta_{p-1}}^{\alpha_{p-1} \beta_{p-1}}
  g^{\sigma \beta} g^{\tau \nu}\phi_{p|\alpha\beta}\gamma_{i\mu} u_\nu \nonumber \\
  &= -\dfrac{1}{2^p}\delta_{\alpha_1 \beta_1 \ldots \alpha_{p-1} \beta_{p-1} \beta \nu}^{\gamma_1 \delta_1 .\ldots \gamma_{p-1} \delta_{p-1} \alpha \mu} \Rr_{\gamma_1 \delta_1}^{\alpha_1 \beta_1} \ldots \Rr_{\gamma_{p-1} \delta_{p-1}}^{\alpha_{p-1} \beta_{p-1}}
  \phi_{p|\alpha}^{\phantom{p}|\beta} \gamma_{i\mu} u^\nu \nonumber \\
  &= -\dfrac{1}{2^p}\delta_{\alpha_1 \beta_1 \ldots \alpha_{p-1} \beta_{p-1} \beta \nu}^{\gamma_1 \delta_1 .\ldots \gamma_{p-1} \delta_{p-1} \alpha \mu}
  \left(\Mr_{\gamma_1 \delta_1}^{\alpha_1 \beta_1} + \Nrb_{\gamma_1 \delta_1}^{\alpha_1 \beta_1} + \Orb_{\gamma_1 \delta_1}^{\alpha_1 \beta_1}\right) \nonumber \\
  & \hspace{15mm }\ldots
  \left(\Mr_{\gamma_{p-1} \delta_{p-1}}^{\alpha_{p-1} \beta_{p-1}} + \Nrb_{\gamma_{p-1} \delta_{p-1}}^{\alpha_{p-1} \beta_{p-1}} + \Orb_{\gamma_{p-1} \delta_{p-1}}^{\alpha_{p-1} \beta_{p-1}}\right)
  \phi_{p|\alpha}^{\phantom{p}|\beta} \gamma_{i\mu} u^\nu.
\end{align}
The only remaining terms are the ones with just one $\Nr_{\gamma}^{\alpha\beta}u_\delta$, for the same reasons as in the precedent paragraph; plus the term with only $\Mr_{\gamma\delta}^{\alpha\beta}$ which does not vanish here:
\begin{align}
P_{(p)}^{\mu\alpha\beta\nu}\phi_{p|\alpha\beta}\gamma_{i\mu} u_\nu
  = - \dfrac{1}{2^p}\bigg[2p &\delta_{\alpha_1 \beta_1 \ldots \alpha_{p-1} \beta_{p-1} \beta \nu}^{\gamma_1 \delta_1 .\ldots \gamma_{p-1} \delta_{p-1} \alpha \mu} \Mr_{\gamma_1 \delta_1}^{\alpha_1 \beta_1} \ldots \Mr_{\gamma_{p-2} \delta_{p-2}}^{\alpha_{p-2} \beta_{p-2}} \Nr_{\gamma_{p-1}}^{\alpha_{p-1} \beta_{p-1}} u_{\delta_{p-1}} \gamma_{i\mu} u^\nu \nonumber \\
  + &\delta_{\alpha_1 \beta_1 \ldots \alpha_{p-1} \beta_{p-1} \beta \nu}^{\gamma_1 \delta_1 .\ldots \gamma_{p-1} \delta_{p-1} \alpha \mu} \Mr_{\gamma_1 \delta_1}^{\alpha_1 \beta_1} \ldots \Mr_{\gamma_{p-2} \delta_{p-2}}^{\alpha_{p-2} \beta_{p-2}} \Mr_{\gamma_{p-1} \delta_{p-1}}^{\alpha_{p-1} \beta_{p-1}} \gamma_{i\mu} u^\nu \bigg] \phi_{p|\alpha}^{\phantom{p}|\beta} \label{Pgu1}
\end{align}

Expanding along the last row, the first term is equal to
\begin{align*}
\delta_{\alpha_1 \beta_1 \ldots \alpha_{p-1} \beta_{p-1} \beta \nu}^{\gamma_1 \delta_1 .\ldots \gamma_{p-1} \delta_{p-1} \alpha \mu}
	& \Mr_{\gamma_1 \delta_1}^{\alpha_1 \beta_1} \ldots \Mr_{\gamma_{p-2} \delta_{p-2}}^{\alpha_{p-2} \beta_{p-2}} \Nr_{\gamma_{p-1}}^{\alpha_{p-1} \beta_{p-1}} u_{\delta_{p-1}} \gamma_{i\mu} u^\nu \\
	=~& \delta_{\alpha_1 \beta_1 \ldots \alpha_{p-1} \beta_{p-1} \beta}^{\gamma_1 \delta_1 .\ldots \gamma_{p-1} \delta_{p-1} \alpha} \delta_\nu^{\mu}
	\Mr_{\gamma_1 \delta_1}^{\alpha_1 \beta_1} \ldots \Mr_{\gamma_{p-2} \delta_{p-2}}^{\alpha_{p-2} \beta_{p-2}} \Nr_{\gamma_{p-1}}^{\alpha_{p-1} \beta_{p-1}} u_{\delta_{p-1}} \gamma_{i\mu} u^\nu \\
	&- \delta_{\alpha_1 \beta_1 \ldots \alpha_{p-1} \beta_{p-1} \beta}^{\gamma_1 \delta_1 .\ldots \gamma_{p-1} \delta_{p-1} \mu} \delta_\nu^{\alpha}
	\Mr_{\gamma_1 \delta_1}^{\alpha_1 \beta_1} \ldots \Mr_{\gamma_{p-2} \delta_{p-2}}^{\alpha_{p-2} \beta_{p-2}} \Nr_{\gamma_{p-1}}^{\alpha_{p-1} \beta_{p-1}} u_{\delta_{p-1}} \gamma_{i\mu} u^\nu \\
	&+ \delta_{\alpha_1 \beta_1 \ldots \alpha_{p-1} \beta_{p-1} \beta}^{\gamma_1 \delta_1 .\ldots \gamma_{p-1} \alpha_{\phantom{p-1}} \mu} \delta_\nu^{\delta_{p-1}}
	\Mr_{\gamma_1 \delta_1}^{\alpha_1 \beta_1} \ldots \Mr_{\gamma_{p-2} \delta_{p-2}}^{\alpha_{p-2} \beta_{p-2}} \Nr_{\gamma_{p-1}}^{\alpha_{p-1} \beta_{p-1}} u_{\delta_{p-1}} \gamma_{i\mu} u^\nu \\
	&- \delta_{\alpha_1 \beta_1 \ldots \alpha_{p-1} \beta_{p-1} \beta}^{\gamma_1 \delta_1 .\ldots \delta_{p-1} \alpha_{\phantom{p-1}} \mu} \delta_\nu^{\gamma_{p-1}}
	\Mr_{\gamma_1 \delta_1}^{\alpha_1 \beta_1} \ldots \Mr_{\gamma_{p-2} \delta_{p-2}}^{\alpha_{p-2} \beta_{p-2}} \Nr_{\gamma_{p-1}}^{\alpha_{p-1} \beta_{p-1}} u_{\delta_{p-1}} \gamma_{i\mu} u^\nu \\
	&+ \delta_{\alpha_1 \beta_1 \ldots \alpha_{p-2} \beta_{p-2} \alpha_{p-1} \beta_{p-1} \beta}^{\gamma_1 \delta_1 .\ldots \gamma_{p-2} \gamma_{p-1} \delta_{p-1} \alpha_{\phantom{p-1}} \mu} \delta_\nu^{\delta_{p-2}}
	\Mr_{\gamma_1 \delta_1}^{\alpha_1 \beta_1} \ldots \Mr_{\gamma_{p-2} \delta_{p-2}}^{\alpha_{p-2} \beta_{p-2}} \Nr_{\gamma_{p-1}}^{\alpha_{p-1} \beta_{p-1}} u_{\delta_{p-1}} \gamma_{i\mu} u^\nu \\
	&- \ldots
\end{align*}
All the lines but the second and the third ones cancel. We permute two indices in the second determinant and expand it along the last row:
\begin{align}
\delta_{\alpha_1 \beta_1 \ldots \alpha_{p-1} \beta_{p-1} \beta \nu}^{\gamma_1 \delta_1 .\ldots \gamma_{p-1} \delta_{p-1} \alpha \mu}
	& \Mr_{\gamma_1 \delta_1}^{\alpha_1 \beta_1} \ldots \Mr_{\gamma_{p-2} \delta_{p-2}}^{\alpha_{p-2} \beta_{p-2}} \Nr_{\gamma_{p-1}}^{\alpha_{p-1} \beta_{p-1}} u_{\delta_{p-1}} \gamma_{i\mu} u^\nu \nonumber \\
	=~& \delta_{\alpha_1 \beta_1 \ldots \alpha_{p-1} \beta_{p-1} \beta}^{\gamma_1 \delta_1 .\ldots \gamma_{p-1} \mu_{\phantom{p-1}} \delta_{p-1}}
	\Mr_{\gamma_1 \delta_1}^{\alpha_1 \beta_1} \ldots \Mr_{\gamma_{p-2} \delta_{p-2}}^{\alpha_{p-2} \beta_{p-2}} \Nr_{\gamma_{p-1}}^{\alpha_{p-1} \beta_{p-1}} u_{\delta_{p-1}} \gamma_{i\mu} u^\alpha \nonumber \\
	&+ \varepsilon\delta_{\alpha_1 \beta_1 \ldots \alpha_{p-1} \beta_{p-1} \beta}^{\gamma_1 \delta_1 .\ldots \gamma_{p-1} \alpha_{\phantom{p-1}} \mu} 
	\Mr_{\gamma_1 \delta_1}^{\alpha_1 \beta_1} \ldots \Mr_{\gamma_{p-2} \delta_{p-2}}^{\alpha_{p-2} \beta_{p-2}} \Nr_{\gamma_{p-1}}^{\alpha_{p-1} \beta_{p-1}} \gamma_{i\mu} \nonumber \\
	=~& \delta_{\alpha_1 \beta_1 \ldots \alpha_{p-1} \beta_{p-1}}^{\gamma_1 \delta_1 .\ldots \gamma_{p-1} \mu} \delta_\beta^{\delta_{p-1}}
	\Mr_{\gamma_1 \delta_1}^{\alpha_1 \beta_1} \ldots \Mr_{\gamma_{p-2} \delta_{p-2}}^{\alpha_{p-2} \beta_{p-2}} \Nr_{\gamma_{p-1}}^{\alpha_{p-1} \beta_{p-1}} u_{\delta_{p-1}} \gamma_{i\mu} u^\alpha \nonumber \\
	&- \varepsilon\delta_{\alpha_1 \beta_1 \ldots \alpha_{p-1} \beta_{p-1} \beta}^{\gamma_1 \delta_1 .\ldots \gamma_{p-1} \mu_{\phantom{p-1}} \alpha}
	\Mr_{\gamma_1 \delta_1}^{\alpha_1 \beta_1} \ldots \Mr_{\gamma_{p-2} \delta_{p-2}}^{\alpha_{p-2} \beta_{p-2}} \Nr_{\gamma_{p-1}}^{\alpha_{p-1} \beta_{p-1}} \gamma_{i\mu} \nonumber \\
	=~& \delta_{\alpha_1 \beta_1 \ldots \alpha_{p-1} \beta_{p-1}}^{\gamma_1 \delta_1 .\ldots \gamma_{p-1} \mu}
	\Mr_{\gamma_1 \delta_1}^{\alpha_1 \beta_1} \ldots \Mr_{\gamma_{p-2} \delta_{p-2}}^{\alpha_{p-2} \beta_{p-2}} \Nr_{\gamma_{p-1}}^{\alpha_{p-1} \beta_{p-1}} \gamma_{i\mu} u^\alpha u_\beta \nonumber \\
	&- \varepsilon \bigg[
	\delta_{\alpha_1 \beta_1 \ldots \alpha_{p-1} \beta_{p-1}}^{\gamma_1 \delta_1 .\ldots \gamma_{p-1} \mu} \delta_\beta^\alpha
	\Mr_{\gamma_1 \delta_1}^{\alpha_1 \beta_1} \ldots \Mr_{\gamma_{p-2} \delta_{p-2}}^{\alpha_{p-2} \beta_{p-2}} \Nr_{\gamma_{p-1}}^{\alpha_{p-1} \beta_{p-1}} \nonumber \\
	&\phantom{- \varepsilon \bigg[}- \delta_{\alpha_1 \beta_1 \ldots \alpha_{p-1} \beta_{p-1}}^{\gamma_1 \delta_1 .\ldots \gamma_{p-1} \alpha} \delta_\beta^\mu
	\Mr_{\gamma_1 \delta_1}^{\alpha_1 \beta_1} \ldots \Mr_{\gamma_{p-2} \delta_{p-2}}^{\alpha_{p-2} \beta_{p-2}} \Nr_{\gamma_{p-1}}^{\alpha_{p-1} \beta_{p-1}} \nonumber \\
	&\phantom{- \varepsilon \bigg[}+ \delta_{\alpha_1 \beta_1 \ldots \alpha_{p-1} \beta_{p-1}}^{\gamma_1 \delta_1 .\ldots \mu_{\phantom{p-1}} \alpha} \delta_\beta^{\gamma_{p-1}}
	\Mr_{\gamma_1 \delta_1}^{\alpha_1 \beta_1} \ldots \Mr_{\gamma_{p-2} \delta_{p-2}}^{\alpha_{p-2} \beta_{p-2}} \Nr_{\gamma_{p-1}}^{\alpha_{p-1} \beta_{p-1}} \nonumber \\
	&\phantom{- \varepsilon \bigg[}-2(p-2) \delta_{\alpha_1 \beta_1 \ldots \alpha_{p-2} \beta_{p-2} \alpha_{p-1} \beta_{p-1}}^{\gamma_1 \delta_1 .\ldots \gamma_{p-2} \gamma_{p-1} \mu_{\phantom{p-1}} \alpha} \delta_\beta^{\delta_{p-2}}
	\Mr_{\gamma_1 \delta_1}^{\alpha_1 \beta_1} \ldots \Mr_{\gamma_{p-2} \delta_{p-2}}^{\alpha_{p-2} \beta_{p-2}} \Nr_{\gamma_{p-1}}^{\alpha_{p-1} \beta_{p-1}} \bigg] \gamma_{i\mu} \nonumber \\
	=~& \delta_{\alpha_1 \beta_1 \ldots \alpha_{p-1} \beta_{p-1}}^{\gamma_1 \delta_1 .\ldots \gamma_{p-1} \mu}
	\Mr_{\gamma_1 \delta_1}^{\alpha_1 \beta_1} \ldots \Mr_{\gamma_{p-2} \delta_{p-2}}^{\alpha_{p-2} \beta_{p-2}} \Nr_{\gamma_{p-1}}^{\alpha_{p-1} \beta_{p-1}} \gamma_{i\mu} \left(u^\alpha u_\beta - \varepsilon \delta_\beta^\alpha\right) \nonumber \\
	&+ \varepsilon \delta_{\alpha_1 \beta_1 \ldots \alpha_{p-1} \beta_{p-1}}^{\gamma_1 \delta_1 .\ldots \gamma_{p-1} \alpha}
	\Mr_{\gamma_1 \delta_1}^{\alpha_1 \beta_1} \ldots \Mr_{\gamma_{p-2} \delta_{p-2}}^{\alpha_{p-2} \beta_{p-2}} \Nr_{\gamma_{p-1}}^{\alpha_{p-1} \beta_{p-1}} \gamma_{i\beta} \nonumber \\
	&- \varepsilon \delta_{\alpha_1 \beta_1 \ldots \alpha_{p-1} \beta_{p-1}}^{\gamma_1 \delta_1 .\ldots \mu_{\phantom{p-1}} \alpha}
	\Mr_{\gamma_1 \delta_1}^{\alpha_1 \beta_1} \ldots \Mr_{\gamma_{p-2} \delta_{p-2}}^{\alpha_{p-2} \beta_{p-2}} \Nr_{\beta}^{\alpha_{p-1} \beta_{p-1}} \gamma_{i\mu} \nonumber \\
	&+ 2 \varepsilon (p-2) \delta_{\alpha_1 \beta_1 \ldots \alpha_{p-1} \beta_{p-1} \alpha_{p-2} \beta_{p-2}}^{\gamma_1 \delta_1 .\ldots \gamma_{p-1} \mu_{\phantom{p-1}} \gamma_{p-2} \alpha}
	\Mr_{\gamma_1 \delta_1}^{\alpha_1 \beta_1} \ldots \Nr_{\gamma_{p-1}}^{\alpha_{p-1} \beta_{p-1}} \Mr_{\gamma_{p-2} \beta}^{\alpha_{p-2} \beta_{p-2}} \gamma_{i\mu} \nonumber \\
	=~& 2^{p-1}\Nb_{(p-1)}^{\phantom{(p-1)}\mu} \gamma_{i\mu} \left(-\varepsilon \gamma_\beta^\alpha\right) + \varepsilon 2^{p-1}\Nb_{(p-1)}^{\phantom{(p-1)}\alpha} \gamma_{i\beta} \nonumber \\
	&- \varepsilon 2^{p-2}\Mbdd_{(p-2)k l}^{\phantom{(p-2)}\mu \alpha} \Nr_\beta^{k l} \gamma_{i\mu} + \varepsilon (p-2) 2^{p-1} \Nbdd_{(p-2)\phantom{\mu} k l}^{\phantom{(p-2)}\mu j \alpha} \Mr_{j \beta}^{k l} \gamma_{i\mu}. \label{mom-first}
\end{align}

Meanwhile, the second term is
\begin{align}
\delta_{\alpha_1 \beta_1 \ldots \alpha_{p-1} \beta_{p-1} \beta \nu}^{\gamma_1 \delta_1 .\ldots \gamma_{p-1} \delta_{p-1} \alpha \mu}
	&\Mr_{\gamma_1 \delta_1}^{\alpha_1 \beta_1} \ldots \Mr_{\gamma_{p-2} \delta_{p-2}}^{\alpha_{p-2} \beta_{p-2}} \Mr_{\gamma_{p-1} \delta_{p-1}}^{\alpha_{p-1} \beta_{p-1}} \gamma_{i\mu} u^\nu \nonumber \\
  &= -\delta_{\alpha_1 \beta_1 \ldots \alpha_{p-1} \beta_{p-1} \beta}^{\gamma_1 \delta_1 .\ldots \gamma_{p-1} \delta_{p-1} \mu} \delta_\nu^\alpha
  \Mr_{\gamma_1 \delta_1}^{\alpha_1 \beta_1} \ldots \Mr_{\gamma_{p-1} \delta_{p-1}}^{\alpha_{p-1} \beta_{p-1}} \gamma_{i\mu} u^{\nu} \nonumber \\
  &= -\delta_{\alpha_1 \beta_1 \ldots \alpha_{p-1} \beta_{p-1} \beta}^{\gamma_1 \delta_1 .\ldots \gamma_{p-1} \delta_{p-1} \mu}
  \Mr_{\gamma_1 \delta_1}^{\alpha_1 \beta_1} \ldots \Mr_{\gamma_{p-1} \delta_{p-1}}^{\alpha_{p-1} \beta_{p-1}}
  \gamma_{i\mu} u^{\alpha} \nonumber \\
  &= -2^{p-1}\Mbd_{(p-1)\beta}^{\phantom{(p-1)}\mu} \gamma_{i\mu} u^{\alpha}. \label{mom-second}
\end{align}

We put \eqref{mom-first} and \eqref{mom-second} together, and obtain
\begin{align*}
P_{(p)}^{\mu\alpha\beta\nu}\phi_{p|\alpha\beta}\gamma_{i\mu} u_\nu
  = \dfrac{1}{2} \bigg[ & \varepsilon p \Big( 2 \Nb_{(p-1)i} \gamma_\beta^\alpha - 2 \Nb_{(p-1)}^{\phantom{(p-1)}\alpha} \gamma_{i\beta}
  + \Mbdd_{(p-2)k l}^{\phantom{(p-2)}\mu \alpha} \Nr_\beta^{k l} \gamma_{i\mu} \\
  &\hspace{10mm} - 2(p-2) \Nbdd_{(p-2)\phantom{\mu} k l}^{\phantom{(p-2)}\mu j \alpha} \Mr_{j \beta}^{k l} \gamma_{i\mu} \Big)
  + \Mbd_{(p-1)\beta}^{\phantom{(p-1)}\mu} \gamma_{i\mu} u^{\alpha} \bigg]\phi_{p|\alpha}^{\phantom{p}|\beta} \\
  = \dfrac{1}{2} \bigg[ & \varepsilon p \Big( 2 \Nb_{(p-1)i} \gamma_b^a - 2 \Nb_{(p-1)}^{\phantom{(p-1)}a} \gamma_{ib}
  + \Mbdd_{(p-2)k l}^{\phantom{(p-2)} m a} \Nr_b^{k l} \gamma_{i m} \\
  &\hspace{10mm} - 2(p-2) \Nbdd_{(p-2)\phantom{m} k l}^{\phantom{(p-2)}m j a} \Mr_{j b}^{k l} \gamma_{im} \Big)\left(\gamma_a^\alpha \gamma_\beta^b \phi_{p|\alpha}^{\phantom{p}|\beta}\right) \\
  &+ \Mbd_{(p-1) b}^{\phantom{(p-1)} m} \gamma_{i m} \left(\gamma_\beta^b u^{\alpha} \phi_{p|\alpha}^{\phantom{p}|\beta} \right)\bigg].
\end{align*}
Using \eqref{ggDDphi} and \eqref{ugDDphi}, we get
\begin{align}
P_{(p)}^{\mu\alpha\beta\nu}\phi_{p|\alpha\beta}\gamma_{i\mu} u_\nu
  = \dfrac{1}{2} \bigg[ & \varepsilon p \Big( 2 \Nb_{(p-1)}^{\phantom{(p-1)}m} \gamma_b^a - 2 \Nb_{(p-1)}^{\phantom{(p-1)}a} \gamma_b^m
  + \Mbdd_{(p-2)k l}^{\phantom{(p-2)} m a} \Nr_b^{k l} \nonumber \\
  & \hspace{10mm} - 2(p-2) \Nbdd_{(p-2)\phantom{m} k l}^{\phantom{(p-2)}m j a} \Mr_{j b}^{k l} \Big) \left(\nabla_a \nabla^b \phi_p + \varepsilon \Pi_p K_a^b \right) \nonumber \\
  & + \Mbd_{(p-1) b}^{\phantom{(p-1)} m} \left(\nabla^b \Pi_p - K^{ab}\nabla_a \phi_p \right)\bigg] \gamma_{im}. \label{Pgu2}
\end{align}

So,
\begin{align}
\Ac_{(p)}^{\mu\nu}\gamma_{i\mu} u_\nu
  =~& \left[\phi_p\Rbd_{(p)}^{\mu\nu} +2p\left(P_{(p)}^{\mu\rho\sigma\nu} + P_{(p)}^{\mu\sigma\rho\nu}\right)\phi_{p|\rho\sigma}\right] \gamma_{i\mu} u_\nu \nonumber \\
	=~& \phi_p\left(-\varepsilon 2p \Nb_{(p)i}\right) \nonumber \\
	& + 2\varepsilon p^2 \bigg( 2 \Nb_{(p-1)}^{\phantom{(p-1)}m} \gamma_b^a - 2 \Nb_{(p-1)}^{\phantom{(p-1)}a} \gamma_b^m \nonumber \\
	& \hspace{15mm} + \Mbdd_{(p-2)k l}^{\phantom{(p-2)} m a} \Nr_b^{k l} - 2(p-2) \Nbdd_{(p-2)\phantom{m} k l}^{\phantom{(p-2)}m j a} \Mr_{j b}^{k l} \bigg) \left(\nabla_a \nabla^b \phi_p + \varepsilon \Pi_p K_a^b \right) \gamma_{im} \nonumber \\
	& + 2p \Mbd_{(p-1) b}^{\phantom{(p-1)} m} \left(\nabla^b \Pi_p - K^{ab}\nabla_a \phi_p \right)\gamma_{im}. \label{Agu}
\end{align}

\subsection*{Dynamical equations}
Similar orthogonality arguments lead to
\begin{align}
\Rbd_{(p)}^{\mu\nu}\gamma_{i\mu} \gamma_{j\nu}
	= \Rbd_{(p)\mu}^{\phantom{(p)}\nu}\gamma_i^\mu \gamma_{j\nu} \nonumber \\
	= \dfrac{1}{2^p} \delta_{\alpha_1 \beta_1 \ldots \alpha_p \beta_p \mu}^{\gamma_1 \delta_1 \ldots . \gamma_p \delta_p \nu}
	&\Rr_{\gamma_1 \delta_1}^{\alpha_1 \beta_1} \Rr_{\gamma_2 \delta_2}^{\alpha_2 \beta_2} \ldots \Rr_{\gamma_p \delta_p}^{\alpha_p \beta_p}\gamma_i^\mu \gamma_{j\nu} \nonumber \\
	= \dfrac{1}{2^p} \delta_{\alpha_1 \beta_1 \ldots \alpha_p \beta_p \mu}^{\gamma_1 \delta_1 \ldots . \gamma_p \delta_p \nu}
	&\left(\Mr_{\gamma_1 \delta_1}^{\alpha_1 \beta_1} + \Nrb_{\gamma_1 \delta_1}^{\alpha_1 \beta_1} + \Orb_{\gamma_1 \delta_1}^{\alpha_1 \beta_1}\right) \nonumber \\
	& \cdots \nonumber \\
	&\left(\Mr_{\gamma_p \delta_p}^{\alpha_p \beta_p} + \Nrb_{\gamma_p \delta_p}^{\alpha_p \beta_p} + \Orb_{\gamma_p \delta_p}^{\alpha_p \beta_p}\right)\gamma_i^\mu \gamma_{j\nu} \nonumber \\
		= \dfrac{1}{2^p} \delta_{\alpha_1 \beta_1 \ldots \alpha_p \beta_p \mu}^{\gamma_1 \delta_1 \ldots . \gamma_p \delta_p \nu}
	&\Big(\Mr_{\gamma_1 \delta_1}^{\alpha_1 \beta_1} \ldots \Mr_{\gamma_p \delta_p}^{\alpha_p \beta_p} \nonumber \\
	& + 4p(p-1) \Mr_{\gamma_1 \delta_1}^{\alpha_1 \beta_1} \ldots \Mr_{\gamma_{p-2} \delta_{p-2}}^{\alpha_{p-2} \beta_{p-2}} \Nr_{\gamma_{p-1} \delta_{p-1}}^{\alpha_{p-1}} u^{\beta_{p-1}} \Nr_{\gamma_p}^{\alpha_p \beta_p} u_{\delta_p} \nonumber \\
	& + 4p \Mr_{\gamma_1 \delta_1}^{\alpha_1 \beta_1} \ldots \Mr_{\gamma_{p-1} \delta_{p-1}}^{\alpha_{p-1} \beta_{p-1}} \Or_{\gamma_{p}}^{\alpha_{p}} u^{\beta_{p}} u_{\delta_{p}} \Big)\gamma_i^\mu \gamma_{j\nu}. \label{Rbdgg1}
\end{align}
The first term is equal to
\begin{align}
\dfrac{1}{2^p} \delta_{\alpha_1 \beta_1 \ldots \alpha_p \beta_p \mu}^{\gamma_1 \delta_1 \ldots . \gamma_p \delta_p \nu} \Mr_{\gamma_1 \delta_1}^{\alpha_1 \beta_1} \ldots \Mr_{\gamma_p \delta_p}^{\alpha_p \beta_p} \gamma_i^\mu \gamma_{j\nu}
	&= \Mbd_{(p)\mu}^{\phantom{(p)}\nu} \gamma_i^\mu \gamma_{j\nu} \nonumber \\
	&= \Mbd_{(p)ij}. \label{dynR-first}
\end{align}
The second term, expanding along a row, is equal to
\begin{align}
\dfrac{4p(p-1)}{2^p}
	&\delta_{\alpha_1 \beta_1 \ldots \alpha_p \beta_p \mu}^{\gamma_1 \delta_1 \ldots . \gamma_p \delta_p \nu} \Mr_{\gamma_1 \delta_1}^{\alpha_1 \beta_1} \ldots \Mr_{\gamma_{p-2} \delta_{p-2}}^{\alpha_{p-2} \beta_{p-2}} \Nr_{\gamma_{p-1} \delta_{p-1}}^{\alpha_{p-1}} u^{\beta_{p-1}} \Nr_{\gamma_p}^{\alpha_p \beta_p} u_{\delta_p} \gamma_i^\mu \gamma_{j\nu} \nonumber \\
	&= \dfrac{4p(p-1)}{2^p} \delta_{\alpha_1 \beta_1 \ldots \alpha_{p-1} \alpha_{p\phantom{-1}} \beta_p \mu}^{\gamma_1 \delta_1 \ldots . \gamma_{p-1} \delta_{p-1} \gamma_p \nu}\delta_{\beta_{p-1}}^{\delta_p} \Mr_{\gamma_1 \delta_1}^{\alpha_1 \beta_1} \ldots \Mr_{\gamma_{p-2} \delta_{p-2}}^{\alpha_{p-2} \beta_{p-2}} \Nr_{\gamma_{p-1} \delta_{p-1}}^{\alpha_{p-1}} u^{\beta_{p-1}} \Nr_{\gamma_p}^{\alpha_p \beta_p} u_{\delta_p} \gamma_i^\mu \gamma_{j\nu} \nonumber \\
	&= \varepsilon \dfrac{4p(p-1)}{2^p} \delta_{\alpha_1 \beta_1 \ldots \alpha_{p-1} \alpha_{p\phantom{-1}} \beta_p \mu}^{\gamma_1 \delta_1 \ldots . \gamma_{p-1} \delta_{p-1} \gamma_p \nu} \Mr_{\gamma_1 \delta_1}^{\alpha_1 \beta_1} \ldots \Mr_{\gamma_{p-2} \delta_{p-2}}^{\alpha_{p-2} \beta_{p-2}} \Nr_{\gamma_{p-1} \delta_{p-1}}^{\alpha_{p-1}} \Nr_{\gamma_p}^{\alpha_p \beta_p} \gamma_i^\mu \gamma_{j\nu} \nonumber \\
	&= \varepsilon \dfrac{4p(p-1)}{2} \NNbd_{(p-1)\mu}^{\phantom{(p-1)}\nu} \gamma_i^\mu \gamma_{j\nu} \nonumber \\
	&= \varepsilon 2p(p-1)\NNbd_{(p-1)ij}. \label{dynR-second}
\end{align}
Finally, the third term is equal to
\begin{align}
\dfrac{4p}{2^p} & \delta_{\alpha_1 \beta_1 \ldots \alpha_p \beta_p \mu}^{\gamma_1 \delta_1 \ldots . \gamma_p \delta_p \nu} \Mr_{\gamma_1 \delta_1}^{\alpha_1 \beta_1} \ldots \Mr_{\gamma_{p-1} \delta_{p-1}}^{\alpha_{p-1} \beta_{p-1}} \Or_{\gamma_{p}}^{\alpha_{p}} u^{\beta_{p}} u_{\delta_{p}} \gamma_i^\mu \gamma_{j\nu} \nonumber \\
	&= \dfrac{4p}{2^p} \delta_{\alpha_1 \beta_1 \ldots \alpha_p \mu}^{\gamma_1 \delta_1 \ldots . \gamma_p \nu} \delta_{\beta_p}^{\delta_p} \Mr_{\gamma_1 \delta_1}^{\alpha_1 \beta_1} \ldots \Mr_{\gamma_{p-1} \delta_{p-1}}^{\alpha_{p-1} \beta_{p-1}} \Or_{\gamma_{p}}^{\alpha_{p}} u^{\beta_{p}} u_{\delta_{p}} \gamma_i^\mu \gamma_{j\nu} \nonumber \\
	&= \varepsilon \dfrac{2p}{2^{p-1}} \delta_{\alpha_1 \beta_1 \ldots \alpha_p \mu}^{\gamma_1 \delta_1 \ldots . \gamma_p \nu} \Mr_{\gamma_1 \delta_1}^{\alpha_1 \beta_1} \ldots \Mr_{\gamma_{p-1} \delta_{p-1}}^{\alpha_{p-1} \beta_{p-1}} \Or_{\gamma_{p}}^{\alpha_{p}} \gamma_i^\mu \gamma_{j\nu} \nonumber \\
	&= \varepsilon 2p \Obd_{(p-1)\mu}^{\phantom{(p-1)}\nu} \gamma_i^\mu \gamma_{j\nu} \nonumber \\
	&= \varepsilon 2p \Obd_{(p-1)ij}. \label{dynR-third}
\end{align}
So \eqref{Rbdgg1} can be written as the sum of \eqref{dynR-first}, \eqref{dynR-second} and \eqref{dynR-third}:
\begin{align}
\Rbd_{(p)}^{\mu\nu} \gamma_{i\mu} \gamma_{j\nu} = \Mbd_{(p)ij} + \varepsilon 2p(p-1)\NNbd_{(p-1)ij} + \varepsilon 2p\Obd_{(p-1)ij}. \label{Rbdgg2}
\end{align}

Now the second projection:
\begin{align}
P_{(p)}^{\mu\alpha\beta\nu} \phi_{p|\alpha\beta} \gamma_{i\mu} \gamma_{j\nu}
	&= \dfrac{1}{2}\Rbdd_{(p-1)\sigma\tau}^{\phantom{(p-1)}\mu\alpha} g^{\sigma\beta} g^{\tau\nu} \phi_{p|\alpha\beta} \gamma_{i\mu} \gamma_{j\nu} \nonumber \\
	&= \dfrac{1}{2^p}\delta_{\alpha_1 \beta_1 \ldots \alpha_{p-1} \beta_{p-1} \sigma \tau}^{\gamma_1 \delta_1 .\ldots \gamma_{p-1} \delta_{p-1} \mu \alpha} \Rr_{\gamma_1 \delta_1}^{\alpha_1 \beta_1} \ldots \Rr_{\gamma_{p-1} \delta_{p-1}}^{\alpha_{p-1} \beta_{p-1}}
  g^{\sigma \beta} g^{\tau \nu}\phi_{p|\alpha\beta}\gamma_{i\mu} \gamma_{j\nu} \nonumber \\
	&= -\dfrac{1}{2^p}\delta_{\alpha_1 \beta_1 \ldots \alpha_{p-1} \beta_{p-1} \nu \beta}^{\gamma_1 \delta_1 .\ldots \gamma_{p-1} \delta_{p-1} \mu \alpha}
	\left( \Mr_{\gamma_1 \delta_1}^{\alpha_1 \beta_1} + \Nrb_{\gamma_1 \delta_1}^{\alpha_1 \beta_1} + \Orb_{\gamma_1 \delta_1}^{\alpha_1 \beta_1}\right) \nonumber \\
	& \hspace{10mm}	\ldots \left( \Mr_{\gamma_{p-1} \delta_{p-1}}^{\alpha_{p-1} \beta_{p-1}} + \Nrb_{\gamma_{p-1} \delta_{p-1}}^{\alpha_{p-1} \beta_{p-1}} + \Orb_{\gamma_{p-1} \delta_{p-1}}^{\alpha_{p-1} \beta_{p-1}} \right)
	\phi_{p|\alpha}^{\phantom{p}|\beta} \gamma_{i\mu} \gamma_j^\nu
\end{align}
which can be written, after cancellation of the orthogonal terms,
\begin{align}
P_{(p)}^{\mu\alpha\beta\nu} \phi_{p|\alpha\beta} \gamma_{i\mu} \gamma_{j\nu} = \hspace{10mm} \nonumber \\
	-\dfrac{1}{2^p}\delta_{\alpha_1 \beta_1 \ldots \alpha_{p-1} \beta_{p-1} \nu \beta}^{\gamma_1 \delta_1 .\ldots \gamma_{p-1} \delta_{p-1} \mu \alpha}
	\bigg(&\Mr_{\gamma_1 \delta_1}^{\alpha_1 \beta_1} \ldots \Mr_{\gamma_{p-2} \delta_{p-2}}^{\alpha_{p-2} \beta_{p-2}} \Mr_{\gamma_{p-1} \delta_{p-1}}^{\alpha_{p-1} \beta_{p-1}} \label{dynP-2-1} \\
	+ 2(p-1) &\Mr_{\gamma_1 \delta_1}^{\alpha_1 \beta_1} \ldots \Mr_{\gamma_{p-2} \delta_{p-2}}^{\alpha_{p-2} \beta_{p-2}} \Nr_{\gamma_{p-1}}^{\alpha_{p-1} \beta_{p-1}} u_{\delta_{p-1}} \label{dynP-2-2} \\
	+ 2(p-1) &\Mr_{\gamma_1 \delta_1}^{\alpha_1 \beta_1} \ldots \Mr_{\gamma_{p-2} \delta_{p-2}}^{\alpha_{p-2} \beta_{p-2}} \Nr_{\gamma_{p-1} \delta_{p-1}}^{\alpha_{p-1}} u^{\beta_{p-1}} \label{dynP-2-3} \\
	+ 4(p-1) &\Mr_{\gamma_1 \delta_1}^{\alpha_1 \beta_1} \ldots \Mr_{\gamma_{p-2} \delta_{p-2}}^{\alpha_{p-2} \beta_{p-2}} \Or_{\gamma_{p-1}}^{\alpha_{p-1}} u^{\beta_{p-1}} u_{\delta_{p-1}} \label{dynP-2-4} \\
	+ 4(p-1)(p-2) &\Mr_{\gamma_1 \delta_1}^{\alpha_1 \beta_1} \ldots \Mr_{\gamma_{p-3} \delta_{p-3}}^{\alpha_{p-3} \beta_{p-3}} \Nr_{\gamma_{p-2} \delta_{p-2}}^{\alpha_{p-2}} u^{\beta_{p-2}} \Nr_{\gamma_{p-1}}^{\alpha_{p-1} \beta_{p-1}} u_{\delta_{p-1}} \bigg) \phi_{p|\alpha}^{\phantom{p}|\beta} \gamma_{i\mu} \gamma_j^\nu. \label{dynP-2-5}
\end{align}
Using the fact that
\begin{align}
\delta_\alpha^\beta
	&= \gamma_\alpha^\beta + \varepsilon u_\alpha u^\beta \nonumber \\
	&= \gamma_\alpha^a \gamma_b^\beta \delta_a^b + \varepsilon u_\alpha u^\beta, \label{trickuu}
\end{align}
we obtain for the first line \eqref{dynP-2-1}
\begin{align}
-\dfrac{1}{2^p}
	&\delta_{\alpha_1 \beta_1 \ldots \alpha_{p-1} \beta_{p-1} \nu \beta}^{\gamma_1 \delta_1 .\ldots \gamma_{p-1} \delta_{p-1} \mu \alpha} \Mr_{\gamma_1 \delta_1}^{\alpha_1 \beta_1} \ldots \Mr_{\gamma_{p-2} \delta_{p-2}}^{\alpha_{p-2} \beta_{p-2}} \Mr_{\gamma_{p-1} \delta_{p-1}}^{\alpha_{p-1} \beta_{p-1}} \phi_{p|\alpha}^{\phantom{p}|\beta} \gamma_{i\mu} \gamma_j^\nu \nonumber \\
	= -\dfrac{1}{2^p}\bigg(
	+ &\delta_{\alpha_1 \beta_1 \ldots \alpha_{p-1} \beta_{p-1} \nu}^{\gamma_1 \delta_1 .\ldots \gamma_{p-1} \delta_{p-1} \mu} \delta_\beta^\alpha \nonumber \\
	- &\delta_{\alpha_1 \beta_1 \ldots \alpha_{p-1} \beta_{p-1} \nu}^{\gamma_1 \delta_1 .\ldots \gamma_{p-1} \delta_{p-1} \alpha} \delta_\beta^\mu \nonumber \\
	+ 2(p-1)&\delta_{\alpha_1 \beta_1 \ldots \alpha_{p-1} \beta_{p-1} \nu}^{\gamma_1 \delta_1 .\ldots \gamma_{p-1} \mu_{\phantom{p-1}} \alpha} \delta_\beta^{\delta_{p-1}} \bigg) \Mr_{\gamma_1 \delta_1}^{\alpha_1 \beta_1} \ldots \Mr_{\gamma_{p-2} \delta_{p-2}}^{\alpha_{p-2} \beta_{p-2}} \Mr_{\gamma_{p-1} \delta_{p-1}}^{\alpha_{p-1} \beta_{p-1}} \phi_{p|\alpha}^{\phantom{p}|\beta} \gamma_{i\mu} \gamma_j^\nu \nonumber \\
	= -\dfrac{1}{2^p}\bigg(
	+ &\delta_{\alpha_1 \beta_1 \ldots \alpha_{p-1} \beta_{p-1} \nu}^{\gamma_1 \delta_1 .\ldots \gamma_{p-1} \delta_{p-1} \mu} \left(\delta_b^a \gamma_a^\alpha \gamma_\beta^b + \varepsilon u_\beta u^\alpha\right) \nonumber \\
	- &\delta_{\alpha_1 \beta_1 \ldots \alpha_{p-1} \beta_{p-1} \nu}^{\gamma_1 \delta_1 .\ldots \gamma_{p-1} \delta_{p-1} a} \delta_b^\mu \gamma_a^\alpha \gamma_\beta^b \nonumber \\
	+ 2(p-1)&\delta_{\alpha_1 \beta_1 \ldots \alpha_{p-1} \beta_{p-1} \nu}^{\gamma_1 \delta_1 .\ldots \gamma_{p-1} \mu_{\phantom{p-1}} a} \delta_b^{\delta_{p-1}} \gamma_a^\alpha \gamma_\beta^b \bigg) \Mr_{\gamma_1 \delta_1}^{\alpha_1 \beta_1} \ldots \Mr_{\gamma_{p-2} \delta_{p-2}}^{\alpha_{p-2} \beta_{p-2}} \Mr_{\gamma_{p-1} \delta_{p-1}}^{\alpha_{p-1} \beta_{p-1}} \phi_{p|\alpha}^{\phantom{p}|\beta} \gamma_{i\mu} \gamma_j^\nu \nonumber \\
	= -\dfrac{1}{2^p}\bigg(
	+ &\delta_{\alpha_1 \beta_1 \ldots \alpha_{p-1} \beta_{p-1} \nu b}^{\gamma_1 \delta_1 .\ldots \gamma_{p-1} \delta_{p-1} \mu a} \gamma_a^\alpha \gamma_\beta^b \nonumber \\
	+ \varepsilon &\delta_{\alpha_1 \beta_1 \ldots \alpha_{p-1} \beta_{p-1} \nu}^{\gamma_1 \delta_1 .\ldots \gamma_{p-1} \delta_{p-1} \mu} u_\beta u^\alpha \bigg) \Mr_{\gamma_1 \delta_1}^{\alpha_1 \beta_1} \ldots \Mr_{\gamma_{p-2} \delta_{p-2}}^{\alpha_{p-2} \beta_{p-2}} \Mr_{\gamma_{p-1} \delta_{p-1}}^{\alpha_{p-1} \beta_{p-1}} \phi_{p|\alpha}^{\phantom{p}|\beta} \gamma_{i\mu} \gamma_j^\nu \nonumber \\
	= -\dfrac{1}{2}\bigg(
	+ & \Mbdd_{(p-1)n b}^{\phantom{(p-1)}m a} \left(\phi_{p|\alpha}^{\phantom{p}|\beta} \gamma_a^\alpha \gamma_\beta^b \right)
	+ \varepsilon \Mbd_{(p-1)n}^{\phantom{(p-1)}m} \left(\phi_{p|\alpha}^{\phantom{p}|\beta} u_\beta u^\alpha \right) \bigg) \gamma_{i m} \gamma_j^n. \label{dynP-2-1bis}
\end{align}
The second line \eqref{dynP-2-2} is
\begin{align}
-\dfrac{2(p-1)}{2^p}
	&\delta_{\alpha_1 \beta_1 \ldots \alpha_{p-1} \beta_{p-1} \beta \nu}^{\gamma_1 \delta_1 .\ldots \gamma_{p-1} \delta_{p-1} \alpha \mu}
	\Mr_{\gamma_1 \delta_1}^{\alpha_1 \beta_1} \ldots \Mr_{\gamma_{p-2} \delta_{p-2}}^{\alpha_{p-2} \beta_{p-2}} \Nr_{\gamma_{p-1}}^{\alpha_{p-1} \beta_{p-1}} u_{\delta_{p-1}} \phi_{p|\alpha}^{\phantom{p}|\beta} \gamma_{i\mu} \gamma_j^\nu \nonumber \\
	= -\dfrac{2(p-1)}{2^p} \bigg(
	+ &\delta_{\alpha_1 \beta_1 \ldots \alpha_{p-1} \beta_{p-1}}^{\gamma_1 \delta_1 .\ldots \gamma_{p-1} \mu} \delta^{\delta_{p-1}}_\beta \delta_\nu^{\alpha}
	\nonumber \\
	- &\delta_{\alpha_1 \beta_1 \ldots \alpha_{p-1} \nu}^{\gamma_1 \delta_1 .\ldots \gamma_{p-1} \mu} \delta^{\delta_{p-1}}_\beta \delta_{\beta_{p-1}}^{\alpha}
	\nonumber \\
	+ &\delta_{\alpha_1 \beta_1 \ldots \beta_{p-1} \nu}^{\gamma_1 \delta_1 .\ldots \gamma_{p-1} \mu} \delta^{\delta_{p-1}}_\beta \delta_{\alpha_{p-1}}^{\alpha}
	\nonumber \\
	-2(p-2) &\delta_{\alpha_1 \beta_1 \ldots \alpha_{p-2} \alpha_{p-1} \beta_{p-1} \nu}^{\gamma_1 \delta_1 .\ldots \gamma_{p-2} \delta_{p-2} \gamma_{p-1} \mu} \delta^{\delta_{p-1}}_\beta \delta_{\beta_{p-2}}^{\alpha} \bigg)
	\Mr_{\gamma_1 \delta_1}^{\alpha_1 \beta_1} \ldots \Mr_{\gamma_{p-2} \delta_{p-2}}^{\alpha_{p-2} \beta_{p-2}} \Nr_{\gamma_{p-1}}^{\alpha_{p-1} \beta_{p-1}} u_{\delta_{p-1}} \phi_{p|\alpha}^{\phantom{p}|\beta} \gamma_{i\mu} \gamma_j^\nu \nonumber \\
	= -\dfrac{2(p-1)}{2^p} \bigg(
	+ & 2^{p-1} \Nb_{(p-1)}^{\phantom{(p-1)}\mu} \delta_\nu^\alpha \nonumber \\
	- & 2^{p-2} \Mbdd_{(p-2) \alpha_{p-1} \nu}^{\phantom{(p-2)} \gamma_{p-1} \mu} \Nr_{\gamma_{p-1}}^{\alpha_{p-1} \alpha} \nonumber \\
	+ & 2^{p-2} \Mbdd_{(p-2) \beta_{p-1} \nu}^{\phantom{(p-2)} \gamma_{p-1} \mu} \Nr_{\gamma_{p-1}}^{\beta_{p-1} \alpha} \nonumber \\
	-2(p-2) \times & 2^{p-2} \Nbdd_{(p-2) \phantom{\mu} \alpha_{p-2} \nu}^{\phantom{(p-2)} \mu \gamma_{p-2} \delta_{p-2}} \Mr_{\gamma_{p-2} \delta_{p-2}}^{\alpha_{p-2} \alpha} \bigg) u_{\beta} \phi_{p|\alpha}^{\phantom{p}|\beta} \gamma_{i\mu} \gamma_j^\nu \nonumber \\
	= (p-1) \bigg(
	- &\Nb_{(p-1)}^{\phantom{(p-1)}m} \delta_n^a
	+ \Mbdd_{(p-2)l n}^{\phantom{(p-2)} k m} \Nr_{k}^{l \, a}
	+ (p-2) \Nbdd_{(p-2) \phantom{m} k n}^{\phantom{(p-2)} m c d} \Mr_{c d}^{k a} \bigg) \left(\gamma_a^\alpha u_{\beta} \phi_{p|\alpha}^{\phantom{p}|\beta}\right) \gamma_{im} \gamma_j^n. \label{dynP-2-2bis}
\end{align}
The third line \eqref{dynP-2-3} is similar:
\begin{align}
-\dfrac{2(p-1)}{2^p}
	&\delta_{\alpha_1 \beta_1 \ldots \alpha_{p-1} \beta_{p-1} \beta \nu}^{\gamma_1 \delta_1 .\ldots \gamma_{p-1} \delta_{p-1} \alpha \mu}
	\Mr_{\gamma_1 \delta_1}^{\alpha_1 \beta_1} \ldots \Mr_{\gamma_{p-2} \delta_{p-2}}^{\alpha_{p-2} \beta_{p-2}} \Nr_{\gamma_{p-1} \delta_{p-1}}^{\alpha_{p-1}} u^{\beta_{p-1}} \phi_{p|\alpha}^{\phantom{p}|\beta} \gamma_{i\mu} \gamma_j^\nu \nonumber \\
	= (p-1) \bigg(
	- &\Nb_{(p-1)n} \delta_b^m
	+ \Mbdd_{(p-2)l n}^{\phantom{(p-2)} k m} \Nr_{k b}^{l}
	+ (p-2) \Nbdd_{(p-2) n c d}^{\phantom{(p-2) n} k m} \Mr_{k b}^{c d} \bigg) \left(\gamma_\beta^b u^\alpha \phi_{p|\alpha}^{\phantom{p}|\beta}\right) \gamma_{im} \gamma_j^n. \label{dynP-2-3bis}
\end{align}
The fourth line \eqref{dynP-2-4} is
\begin{align*}
-\dfrac{4(p-1)}{2^p}
	&\delta_{\alpha_1 \beta_1 \ldots \alpha_{p-1} \beta_{p-1} \beta \nu}^{\gamma_1 \delta_1 .\ldots \gamma_{p-1} \delta_{p-1} \alpha \mu}
	\Mr_{\gamma_1 \delta_1}^{\alpha_1 \beta_1} \ldots \Mr_{\gamma_{p-2} \delta_{p-2}}^{\alpha_{p-2} \beta_{p-2}} \Or_{\gamma_{p-1}}^{\alpha_{p-1}} u^{\beta_{p-1}} u_{\delta_{p-1}} \phi_{p|\alpha}^{\phantom{p}|\beta} \gamma_{i\mu} \gamma_j^\nu \\
	= -\dfrac{4(p-1)}{2^p} \bigg(
	+ &\delta_{\alpha_1 \beta_1 \ldots \alpha_{p-1} \beta \nu}^{\gamma_1 \delta_1 .\ldots \gamma_{p-1} \alpha \mu} \delta_{\beta_{p-1}}^{\delta_{p-1}} \\
	- &\delta_{\alpha_1 \beta_1 \ldots \alpha_{p-1} \nu}^{\gamma_1 \delta_1 .\ldots \gamma_{p-1} \mu} \delta_{\beta_{p-1}}^{\alpha} \delta_{\beta}^{\delta_{p-1}} \bigg)
	\Mr_{\gamma_1 \delta_1}^{\alpha_1 \beta_1} \ldots \Mr_{\gamma_{p-2} \delta_{p-2}}^{\alpha_{p-2} \beta_{p-2}} \Or_{\gamma_{p-1}}^{\alpha_{p-1}} u^{\beta_{p-1}} u_{\delta_{p-1}} \phi_{p|\alpha}^{\phantom{p}|\beta} \gamma_{i\mu} \gamma_j^\nu \\
	= -\dfrac{4(p-1)}{2^p} \bigg(
	+ & \varepsilon 2^{p-2} \Obdd_{(p-2) \beta \nu}^{\phantom{(p-2)} \alpha \mu}
	- 2^{p-2} \Obd_{(p-2)\nu}^{\phantom{(p-2)}\mu} u^\alpha u_\beta \bigg) \phi_{p|\alpha}^{\phantom{p}|\beta} \gamma_{i\mu} \gamma_j^\nu.
\end{align*}
There we use \eqref{trickuu} again:
\begin{align*}
2^{p-2} \Obdd_{(p-2) \beta \nu}^{\phantom{(p-2)} \alpha \mu} \gamma_{i \mu} \gamma_j^\nu
	&= \delta_{\alpha_1 \beta_1 \ldots \alpha_{p-1} \nu \beta}^{\gamma_1 \delta_1 .\ldots \gamma_{p-1} \mu \alpha}
	\Mr_{\gamma_1 \delta_1}^{\alpha_1 \beta_1} \ldots \Mr_{\gamma_{p-2} \delta_{p-2}}^{\alpha_{p-2} \beta_{p-2}} \Or_{\gamma_{p-1}}^{\alpha_{p-1}} \gamma_{i \mu} \gamma_j^\nu \\
	&= + \delta_{\alpha_1 \beta_1 \ldots \alpha_{p-1} \nu}^{\gamma_1 \delta_1 .\ldots \gamma_{p-1} \mu} \delta_\beta^\alpha
	\Mr_{\gamma_1 \delta_1}^{\alpha_1 \beta_1} \ldots \Mr_{\gamma_{p-2} \delta_{p-2}}^{\alpha_{p-2} \beta_{p-2}} \Or_{\gamma_{p-1}}^{\alpha_{p-1}} \gamma_{i \mu} \gamma_j^\nu \\
	&+ \ldots \\
	&= + \delta_{\alpha_1 \beta_1 \ldots \alpha_{p-1} \nu}^{\gamma_1 \delta_1 .\ldots \gamma_{p-1} \mu} \left( \gamma^\alpha_a \gamma^b_\beta \delta^a_b + \varepsilon u_\beta u^\alpha \right)
	\Mr_{\gamma_1 \delta_1}^{\alpha_1 \beta_1} \ldots \Mr_{\gamma_{p-2} \delta_{p-2}}^{\alpha_{p-2} \beta_{p-2}} \Or_{\gamma_{p-1}}^{\alpha_{p-1}} \gamma_{i \mu} \gamma_j^\nu \\
	&+ \ldots \\
	&= 2^{p-2}\left(\Obdd_{(p-2)\nu b}^{\phantom{(p-2)}\mu a} \gamma^\alpha_a \gamma^b_\beta + \varepsilon \Obd_{(p-2)\nu}^{\phantom{(p-2)}\mu} u_\beta u^\alpha \right)\gamma_{i \mu} \gamma_j^\nu,
\end{align*}
and the fourth line \eqref{dynP-2-4} becomes
\begin{align}
-\dfrac{4(p-1)}{2^p}
	&\delta_{\alpha_1 \beta_1 \ldots \alpha_{p-1} \beta_{p-1} \beta \nu}^{\gamma_1 \delta_1 .\ldots \gamma_{p-1} \delta_{p-1} \alpha \mu}
	\Mr_{\gamma_1 \delta_1}^{\alpha_1 \beta_1} \ldots \Mr_{\gamma_{p-2} \delta_{p-2}}^{\alpha_{p-2} \beta_{p-2}} \Or_{\gamma_{p-1}}^{\alpha_{p-1}} u^{\beta_{p-1}} u_{\delta_{p-1}} \phi_{p|\alpha}^{\phantom{p}|\beta} \gamma_{i\mu} \gamma_j^\nu \nonumber \\
	&= - \varepsilon(p-1) \Obdd_{(p-2) b n}^{\phantom{(p-2)} a m}
	\left(\phi_{p|\alpha}^{\phantom{p}|\beta} \gamma_a^\alpha \gamma_\beta^b \right) \gamma_{i m} \gamma_j^n. \label{dynP-2-4bis}
\end{align}
Finally, to the fifth line \eqref{dynP-2-5} can be used the same calculus:
\begin{align}
	-\dfrac{4(p-1)(p-2)}{2^p}
	&\delta_{\alpha_1 \beta_1 \ldots \alpha_{p-1} \beta_{p-1} \beta \nu}^{\gamma_1 \delta_1 .\ldots \gamma_{p-1} \delta_{p-1} \alpha \mu}
	\Mr_{\gamma_1 \delta_1}^{\alpha_1 \beta_1} \ldots \Mr_{\gamma_{p-3} \delta_{p-3}}^{\alpha_{p-3} \beta_{p-3}} \nonumber \\
	& \phantom{\delta_{\alpha_1 \beta_1 \ldots \alpha_{p-1} \beta_{p-1} \beta \nu}^{\gamma_1 \delta_1 .\ldots \gamma_{p-1} \delta_{p-1} \alpha \mu}}
	\Nr_{\gamma_{p-2} \delta_{p-2}}^{\alpha_{p-2}} u^{\beta_{p-2}} \Nr_{\gamma_{p-1}}^{\alpha_{p-1} \beta_{p-1}} u_{\delta_{p-1}} \phi_{p|\alpha}^{\phantom{p}|\beta} \gamma_{i\mu} \gamma_j^\nu \nonumber \\
	= -\dfrac{4(p-1)(p-2)}{2^p} \bigg(
	+ & \delta_{\alpha_1 \beta_1 \ldots \alpha_{p-2} \alpha_{p-1} \beta_{p-1} \beta \nu}^{\gamma_1 \delta_1 .\ldots \gamma_{p-2} \delta_{p-2} \gamma_{p-1} \alpha \mu} \delta_{\beta_{p-2}}^{\delta_{p-1}} \nonumber \\
	- & \delta_{\alpha_1 \beta_1 \ldots \alpha_{p-2} \alpha_{p-1} \beta_{p-1} \nu}^{\gamma_1 \delta_1 .\ldots \gamma_{p-2} \delta_{p-2} \gamma_{p-1} \mu} \delta_{\beta}^{\delta_{p-1}} \delta_{\beta_{p-2}}^{\alpha} \bigg)
	\Mr_{\gamma_1 \delta_1}^{\alpha_1 \beta_1} \ldots \Mr_{\gamma_{p-3} \delta_{p-3}}^{\alpha_{p-3} \beta_{p-3}} \nonumber \\
	& \phantom{\delta_{\alpha_1 \beta_1 \ldots \alpha_{p-2} \alpha_{p-1} \beta_{p-1} \nu}^{\gamma_1 \delta_1 .\ldots \gamma_{p-2} \delta_{p-2} \gamma_{p-1} \mu} \delta_{\beta}^{\delta_{p-1}} \delta_{\beta_{p-2}}^{\alpha} \bigg)} \Nr_{\gamma_{p-2} \delta_{p-2}}^{\alpha_{p-2}} u^{\beta_{p-2}} \Nr_{\gamma_{p-1}}^{\alpha_{p-1} \beta_{p-1}} u_{\delta_{p-1}} \phi_{p|\alpha}^{\phantom{p}|\beta} \gamma_{i\mu} \gamma_j^\nu \nonumber \\
	= -\dfrac{4(p-1)(p-2)}{2^p} \bigg(
	+ & \varepsilon 2^{p-2} \NNbdd_{(p-2)\beta\nu}^{\phantom{(p-2)}\alpha\mu}
	- 2^{p-2} \NNbd_{(p-2)\nu}^{\phantom{(p-2)}\mu} u_\beta u^\alpha \bigg) \phi_{p|\alpha}^{\phantom{p}|\beta} \gamma_{i\mu} \gamma_j^\nu \nonumber \\
	= - \varepsilon(p-1)(p-2) & \NNbdd_{(p-2)b n}^{\phantom{(p-2)}a m} \left(\phi_{p|\alpha}^{\phantom{p}|\beta}\gamma_a^\alpha \gamma_\beta^b\right) \gamma_{i m} \gamma_j^n. \label{dynP-2-5bis}
\end{align}
We put \eqref{dynP-2-1bis}, \eqref{dynP-2-2bis}, \eqref{dynP-2-3bis}, \eqref{dynP-2-4bis} and \eqref{dynP-2-5bis} together, and obtain
\begin{align}
P_{(p)}^{\mu\alpha\beta\nu} \phi_{p|\alpha\beta} \gamma_{i\mu} \gamma_{j\nu} & = \nonumber \\
	= -\dfrac{1}{2}\bigg(
	+ & \Mbdd_{(p-1)n b}^{\phantom{(p-1)}m a} \left(\phi_{p|\alpha}^{\phantom{p}|\beta} \gamma_a^\alpha \gamma_\beta^b \right)
	+ \varepsilon \Mbd_{(p-1)n}^{\phantom{(p-1)}m} \left(\phi_{p|\alpha}^{\phantom{p}|\beta} u_\beta u^\alpha \right) \bigg) \gamma_{i m} \gamma_j^n \\
	+ (p-1) \bigg(
	- &\Nb_{(p-1)}^{\phantom{(p-1)}m} \delta_n^a
	+ \Mbdd_{(p-2)l n}^{\phantom{(p-2)} k m} \Nr_{k}^{l \, a}
	+ (p-2) \Nbdd_{(p-2) \phantom{m} k n}^{\phantom{(p-2)} m c d} \Mr_{c d}^{k a} \bigg) \left(\gamma_a^\alpha u_{\beta} \phi_{p|\alpha}^{\phantom{p}|\beta}\right) \gamma_{im} \gamma_j^n \nonumber \\
	+ (p-1) \bigg(
	- &\Nb_{(p-1)n} \delta_b^m
	+ \Mbdd_{(p-2)l n}^{\phantom{(p-2)} k m} \Nr_{k b}^{l}
	+ (p-2) \Nbdd_{(p-2) n c d}^{\phantom{(p-2) n} k m} \Mr_{k b}^{c d} \bigg) \left(\gamma_\beta^b u^\alpha \phi_{p|\alpha}^{\phantom{p}|\beta}\right) \gamma_{im} \gamma_j^n \nonumber \\
	- \varepsilon (p-1) \bigg(
	& \Obdd_{(p-2) b n}^{\phantom{(p-2)} a m} + (p-2)\NNbdd_{(p-2) b n}^{\phantom{(p-2)} a m} \bigg)
	 \left(\gamma_a^\alpha \gamma_\beta^b \phi_{p|\alpha}^{\phantom{p}|\beta}\right) \gamma_{i m} \gamma_j^n \nonumber \\
	= -\dfrac{1}{2}\bigg(
	+ & \Mbdd_{(p-1)b n}^{\phantom{(p-1)} a m} + 2\varepsilon (p-1) \left[
	\Obdd_{(p-2) b n}^{\phantom{(p-2)} a m} + (p-2)\NNbdd_{(p-2) b n}^{\phantom{(p-2)} a m}\right] \bigg)
	\left(\nabla_a \nabla^b \phi_p + \varepsilon \Pi_p K_a^b \right) \gamma_{i m} \gamma_j^n \\
	+ (p-1) \bigg(
	- &\Nb_{(p-1)}^{\phantom{(p-1)}m} \delta_n^a
	+ \Mbdd_{(p-2)l n}^{\phantom{(p-2)} k m} \Nr_{k}^{l \, a}
	+ (p-2) \Nbdd_{(p-2) \phantom{m} k n}^{\phantom{(p-2)} m c d} \Mr_{c d}^{k a} \bigg) \left(\nabla_a \Pi_p - K_a^b \nabla_b \phi_p\right) \gamma_{im} \gamma_j^n \nonumber \\
	+ (p-1) \bigg(
	- &\Nb_{(p-1)n} \delta_b^m
	+ \Mbdd_{(p-2)l n}^{\phantom{(p-2)} k m} \Nr_{k b}^{l}
	+ (p-2) \Nbdd_{(p-2) n c d}^{\phantom{(p-2) n} k m} \Mr_{k b}^{c d} \bigg) \left(\nabla^b \Pi_p - K_a^b \nabla^a \phi_p\right) \gamma_{im} \gamma_j^n. \nonumber \\
	-\dfrac{1}{2} & \varepsilon \Mbd_{(p-1)n}^{\phantom{(p-1)}m} \big(\varepsilon\left(\square \phi_p - \Delta \phi_p\right) - \Pi_p K \big) \gamma_{i m} \gamma_j^n. \label{Pgg}
\end{align}

So,
\begin{align}
	&\hspace{20mm} \Ac_{(p)}^{\mu\nu}\gamma_{i\mu} \gamma_{j\nu}
	= \left[\phi_p\Rbd_{(p)}^{\mu\nu} +2p\left(P_{(p)}^{\mu\rho\sigma\nu} + P_{(p)}^{\mu\sigma\rho\nu}\right)\phi_{p|\rho\sigma}\right] \gamma_{i\mu} \gamma_{j\nu} \nonumber \\
	&= \bigg(\Mbd_{(p)ij} + \varepsilon 2p(p-1)\NNbd_{(p-1)ij} + \varepsilon 2p\Obd_{(p-1)ij}\bigg)\phi_p \nonumber \\
	&-p \bigg(\Mbdd_{(p-1)b n}^{\phantom{(p-1)} a m} + 2\varepsilon (p-1)(p-2) \NNbdd_{(p-2) b n}^{\phantom{(p-2)} a m} + 2\varepsilon (p-1)\Obdd_{(p-2) b n}^{\phantom{(p-2)} a m} \bigg)
	\left(\nabla_a \nabla^b \phi_p + \varepsilon \Pi_p K_a^b \right) \gamma_{i m} \gamma_j^n \nonumber \\
	&+ 2p(p-1) \bigg(
	- \Nb_{(p-1)}^{\phantom{(p-1)}m} \delta_n^a
	+ \Mbdd_{(p-2)l n}^{\phantom{(p-2)} k m} \Nr_{k}^{l \, a}
	+ (p-2) \Nbdd_{(p-2) \phantom{m} k n}^{\phantom{(p-2)} m c d} \Mr_{c d}^{k a} \bigg) \left(\nabla_a \Pi_p - K_a^b \nabla_b \phi_p\right) \gamma_{im} \gamma_j^n \nonumber \\
	&+ 2p(p-1) \bigg(
	- \Nb_{(p-1)n} \delta_b^m
	+ \Mbdd_{(p-2)l n}^{\phantom{(p-2)} k m} \Nr_{k b}^{l}
	+ (p-2) \Nbdd_{(p-2) n c d}^{\phantom{(p-2) n} k m} \Mr_{k b}^{c d} \bigg) \left(\nabla^b \Pi_p - K_a^b \nabla^a \phi_p\right) \gamma_{im} \gamma_j^n \nonumber \\
	&-\varepsilon p \Mbd_{(p-1)n}^{\phantom{(p-1)}m} \big(\varepsilon\left(\square \phi_p - \Delta \phi_p\right) - \Pi_p K \big) \gamma_{i m} \gamma_j^n. \label{Agg}
\end{align}

\bibliographystyle{plain}
\bibliography{../../Biblio}

\end{document}